\documentclass[referee,a4paper,12pt,traditabstract]{swsc} 

\usepackage{txfonts}
\usepackage{epstopdf}
\usepackage{lineno}

\usepackage{microtype}						
\usepackage{booktabs}						
\usepackage{tabularx}							
\usepackage{longtable}						
\usepackage{graphicx}						
\usepackage{caption,subcaption}				
\usepackage{listings}							
\usepackage[font=small]{quoting}				
\usepackage[english]{varioref}					
\usepackage{mparhack,relsize}					
\usepackage[usenames,dvipsnames]{xcolor}

\usepackage{hyperref}
\hypersetup{colorlinks=true,citecolor=cyan,urlcolor=cyan,linkcolor=blue}

\usepackage{url}

\usepackage[english]{babel}
\usepackage[usenames,dvipsnames]{xcolor}
\usepackage{bm}

\usepackage{siunitx}							
\DeclareSIUnit\gauss{G}
\DeclareSIUnit\kms{\si{\km \cdot \s^{-1}}}		
\DeclareSIUnit\kmss{\si{\km \cdot \s^{-2}}}		

\usepackage{tikz}
\usetikzlibrary{positioning,matrix,calc,arrows,arrows.meta,decorations.markings,patterns,decorations.pathreplacing, shapes}
\usepackage[all,knot,poly,cmtip]{xy}

\usepackage[style=authortitle, citestyle=authoryear-comp, hyperref,  backend=bibtex, natbib=true, firstinits=true, dashed=false, uniquename=init, maxnames=2]{biblatex}
\DeclareFieldFormat[article]{number}{\mkbibparens{#1}}
\DeclareFieldFormat[incollection]{title}{\mkbibitalic{#1}}
\DeclareNameAlias{sortname}{last-first}
\DeclareNameAlias{default}{last-first}
\DeclareFieldFormat*{journaltitle}{{#1}}

\newcommand{\mockalph}[1]{}

\usepackage[babel]{csquotes}
\usepackage{textcomp}

\bibliography{Bibliography}

\DeclareDelimFormat[cbx@textcite]{nameyeardelim}{\addspace}

\definecolor{webgreen}{rgb}{0,.5,0}
\definecolor{webbrown}{rgb}{.6,0,0}
\definecolor{cobalt}{rgb}{0.0, 0.28, 0.67}

\tikzstyle{title} = [rectangle,						fill=white!20, 			text width=10em, 		text centered, 		rounded corners, 	minimum height=2.3em]
\tikzstyle{block} = [rectangle, draw, 				fill=white!20, 			text width=10em, 		text centered, 		rounded corners, 	minimum height=2.3em]
\tikzstyle{output} = [draw=cobalt,					fill=cobalt!10,  			text width=10em, 		text centered, 		rounded corners, 	minimum height=2.3em]
\tikzstyle{model} = [draw=Turquoise,				fill=Turquoise!10,  		text width=10em, 		text centered, 		rounded corners, 	minimum height=2.3em]
\tikzstyle{cloud} = [draw=red, 	ellipse, 	dashed,	fill=none, 				node distance=2.5cm,		minimum height=2.3em]
\tikzstyle{decision} = [diamond, 					draw, fill=white!20, 		text width=4.5em, 			text badly centered, 	node distance=2.5cm, 	inner sep=0pt]
\tikzstyle{comment} = [								fill=white!20, 			node distance=2.5cm,		minimum height=2em]
\tikzstyle{line} = [draw, ->]

\begin{document}


   \title{Halo Coronal Mass Ejections
   during Solar Cycle 24: reconstruction of the global scenario and geoeffectiveness}
   \titlerunning{Geoeffective CMEs in Solar Cycle 24}
   \authorrunning{C. Scolini et al.}
   \author{C. Scolini
          \inst{1,2}\fnmsep\thanks{corresponding author},     
          M. Messerotti
          \inst{3,4}
          \and
          S. Poedts
          \inst{1}
          \and
          L. Rodriguez
          \inst{2}
          }

   \institute{Centre for mathematical Plasma Astrophysics (CmPA), KU Leuven, 3001 Leuven, Belgium
   	  \and
             Solar-Terrestrial Center of Excellence, SIDC, Royal Observatory of Belgium, 1180 Brussels, Belgium \\
              \email{\href{mailto:camilla.scolini@kuleuven.be}{camilla.scolini@kuleuven.be}}
         \and
             Department of Physics, University of Trieste, 34127 Trieste, Italy
         \and
             INAF-Astronomical Observatory of Trieste, 34149 Trieste, Italy
             }


 
\abstract
{Coronal Mass Ejections (CMEs), in particular Earth-directed ones, are regarded as the main drivers of geomagnetic activity. 
In this study, we present a statistical analysis of a set of 53 fast ($V~\ge~1000$~\si{\kms}) Earth-directed halo CMEs 
observed by the SOHO/LASCO instrument during the period Jan. 2009-Sep. 2015, 
and we then use this CME sample to test the forecasting capabilities of a new Sun-to-Earth prediction scheme for the geoeffectiveness of Earth-directed halo CMEs.
First, we investigate the CME association with other solar activity features such as solar flares, active regions, and others, by means of multi-instrument observations of the solar magnetic and plasma properties, with the final aim of identifying recurrent peculiar features that can be used as precursors of CME-driven geomagnetic storms.
Second, using coronagraphic images to derive the CME kinematical properties at 0.1 AU, 
we propagate the events to 1 AU by means of 3D global MHD simulations. 
In particular, we use the WSA-ENLIL+Cone model to reconstruct the propagation and global evolution of each event up to their arrival at Earth, 
where simulation results are compared with Interplanetary CME (ICME) in-situ signatures.
We then use simulation outputs upstream of Earth to predict their impact on geospace.
By applying the pressure balance condition at the magnetopause and the coupling function proposed by Newell et al. (2008)
to link upstream solar wind properties to the global $K_p$ index,
we estimate the expected magnetospheric compression and geomagnetic activity level, 
and compare our predictions with global data records.

The analysis indicates that 82\% of the fast Earth-directed halo CMEs arrived at Earth within the next 4 days. 
Almost the totality of them compressed the magnetopause below geosynchronous orbits and triggered a minor or major geomagnetic storm afterwards.
Among them, complex sunspot-rich active regions associated with X- and M-class flares are the most favourable configurations from which geoeffective CMEs originate.
The analysis of related Solar Energetic Particle (SEP) events shows that 74\% of the CMEs associated with major SEPs were geoeffective, 
\emph{i.e.} they triggered a minor to intense geomagnetic storm ($K_p \ge 5$). Moreover, the SEP production is enhanced in the case of fast and interacting CMEs.
In this work we present a first attempt at applying a Sun-to-Earth geoeffectiveness prediction scheme - based on 3D simulations and solar wind-geomagnetic activity coupling functions - to a statistical set of fast Earth-directed, potentially geoeffective halo CMEs.
The results of the prediction scheme are promising and in good agreement with the actual data records for geomagnetic activity. However, we point out the need for future studies performing a fine-tuning of the prediction scheme, in particular in terms of the evaluation of the CME input parameters and the modelling of their internal magnetic structure.
}      
   
\keywords{	Coronal Mass Ejections (CMEs) --	
		   	geomagnetic storms --                 	
		   	geoeffectiveness predictions --
                 	Solar Cycle 24 
               }

\maketitle

\section{Introduction}

Coronal Mass Ejections (CMEs) are large-scale eruptions of magnetised plasma from the Sun, and are considered by the space physics community to be the
main drivers of space weather (\cite{gosling:91}; \cite{gosling:93}; \cite{koskinen:06}).
The phenomenon has been defined by \citet{hundhausen:84} as an observable change in coronal structure that: 
(1) occurs on a time scale between a few minutes and several hours and 
(2) involves the appearance and outwards motion of a new, discrete, bright, white-light feature in the coronagraph field of view. 
CMEs usually originate in active regions (ARs) and appear in association with other solar activity signatures, mostly solar flares and filament activations/eruptions.
It is now known that CMEs are extremely common events and occur at a rate which is highly dependent on the solar activity cycle. 
During solar minima, the average rate is $\sim 1$ CME per day, while during the maxima of solar activity, it can exceed the value of 10 CMEs per day 
({\cite{yashiro:04}; \cite{robbrecht:09}).
Among them, Earth-directed, fast ($V \ge 1,000$ \si{\kms}) CMEs observed as halo events in coronagraphs along the Sun-Earth line are by far the most important CME class in terms of space weather implications and effects on Earth (\cite{webb:00}; \cite{michalek:06}). 
They constitute 3\% of all CMEs and tend to originate close to the disk centre, even though about 10\% of them originate close to the limb \citep{gopal:15a}. 
Considering a 73-month period in Solar Cycle 24, one of the weakest cycles ever recorded, a rate of 3.56 halo CMEs per month has been reported \citep{gopal:15b}.
While halo CMEs are the most important ones in terms of impact on geospace, the determination of the kinematical and geometrical properties for such events is particularly difficult when using single-spacecraft observations only, due to the severe projection effects.
Several approaches to cope with such limitations, based either on empirical relations derived from statistical analyses of limb (\emph{e.g.} side-viewed) CME events observed from Earth \citep{gopal:09b}, or on 3D-reconstruction methods based on multi-spacecraft observations (\citet{mierla:10} and references therein), have been proposed. 
Statistical studies show that the average CME width (obtained from non-halo CMEs only) in the rising phase of Solar Cycle 24 is $\sim 55^{\circ}$, 
while the average CME speed is $\sim 650$ \si{\kms} \citep{gopal:14}.

When observed in-situ, the interplanetary (IP) counterparts of CMEs are denoted as Interplanetary CMEs (ICMEs).
They are observed passing over Earth at an average rate of 1-2 per month \citep{richardson:10}.
Several definitions have been proposed in literature to identify and classify the variety of ICME signatures observed from in-situ measurements, 
making the interpretation of the original term somehow ambiguous and uncertain in its meaning.  
Typical in-situ ICME signatures are (\citet{wimmer:06} and references therein):
(1) the presence of a forward shock followed by a turbulent region of highly distorted magnetic field resulting from the compression, deflection and heating of the ambient solar wind, known as sheath region;
(2) enhanced He, O and Fe charged states;
(3) different elemental composition compared to the surrounding solar wind (He/H and Mg/O ratios);
(4) isotopic anomalies ($^{3}$He$^{2+}$/$^{4}$He$^{2+}$);
(5) bidirectional electron streaming;
(6) low proton temperature;
(7) a magnetic structure classified as Magnetic Cloud (MC) \citep{burlaga:91}.
However, none of the above signatures or combination of them gives a foolproof ICME identifier \citep{wimmer:06}. 
For the sake of clarity and as we will deal with fast CMEs only, in the following analysis we will discuss the CME arrival at Earth position in terms of the ICME-driven forward shock.

Impacting CMEs affect technological systems and human activities in several ways \citep{cannon:13}. 
The magnitude of a CME effect on geospace is strongly related to solar wind macroscopic parameters and magnetic properties, 
in particular the interplanetary magnetic field (IMF) vector $\bm{B}$ and the solar wind number density $N$ and bulk speed ${V}$.
While the density and speed determine the compression of the magnetosphere induced by the impinging solar wind, 
a southward magnetic field orientation is mainly responsible for facilitating the dayside magnetopause reconnection and the development of strong disturbances in the geomagnetic field referred to as magnetic storms (\citealt{gonzalez:94}; \citealt{gopal:07}; \citealt{lugaz:15}; \citealt{lugaz:16}).
%
%
Over the decades, several global geomagnetic activity indices have been defined in order to quantify the magnitude of geomagnetic storms recorded at ground level.
Among them, the global Dst and $K_p$ indices are both based on the measured variation of the horizontal component of the on-ground magnetic field as proxy of the geomagnetic activity level, and have been extensively used to evaluate the level of perturbation in the magnetospheric/ionospheric environment.
The Dst index provides a measure of the strength of the equatorial ring current and it is used as reference for the classification of geomagnetic storms \citep{gonzalez:94}.
On the other hand, $K_p$ is a mid-latitude index sensitive to contributions from both auroral and equatorial currents, 
which is used to set the various levels of alert in the NOAA Space Weather Scale for geomagnetic storms (\url{http://www.swpc.noaa.gov/noaa-scales-explanation}).
The two indices are not related by a one-to-one correspondence, as they are sensitive to different current systems developing in the ionosphere and inner magnetosphere during geomagnetic storms (\citealt{kivelson:russell}; \citealt{huttunen:02}; \citealt{huttunen:04}).

Previous studies about the solar and interplanetary sources of geomagnetic storms revealed that about 30\% of all storms (Dst $\le -30$ \si{\nano\tesla}) originate from ICMEs \parencite{zhang:04}.
However, considering intense storms only (Dst $\le -100$ \si{\nano\tesla}, \cite{gonzalez:94}) the fraction reaches up to $\sim$90\% of the total (\cite{zhang:04}; \cite{zhang:07}; see also \citet{richardson:12} for an analysis of the $K_p$ index).
Major disturbances are mostly caused by CMEs coming from the solar disk centre, although some moderate disturbances can be caused by events originated near or at the west limb (\cite{huttunen:02}; \cite{rodriguez:09}; \cite{cid:12}). 
On the other hand, CMEs originated at the disk centre can also be deflected away from the Sun-Earth line 
due to non-radial channelling caused by fast solar wind streams generated by adjacent coronal holes.
\citet{moestl:15} reported the case of a CME - occurred on 7 Jan. 2014 and included in our list as CME \#37 (see Table~\ref{tab:events}) - 
which, despite being originated at the disk centre, was longitudinally deflected to the west due to a coronal hole on the east side of the source region, 
resulting in a weaker impact at Earth than expected.
Previous studies have also found that ICME sheaths alone, \emph{i.e.} not followed by any ICME ejecta, can also be source of intense storms (\cite{tsurutani:88}; \cite{huttunen:02}). The role of CME interactions as sources of strong geomagnetic storms has been investigated by \citet{lugaz:14} and \citet{lugaz:17}.

Another kind of CME-related phenomenon that is of interest for the space weather community is that of Solar Energetic Particles (SEPs).
They mainly consist of protons and electrons that are accelerated to quasi-relativistic speeds.
Large gradual events triggered by CMEs are of particular interest to space travel as they can constitute a significant radiation hazard for astronauts and equipment, 
especially beyond the Earth's magnetic field (\cite{reames:13} and references therein). 
The SEP production is believed to take place at CME forward shocks, 
or in relation to CME-CME interaction phenomena (\cite{cane:06}; \cite{reames:13}; \cite{gopal:15c}).
Concerning the role of fast halo CME as major source of strong SEP events, \citet{gopal:15c} analysed the association of 37 major SEP events in the years 2010-2014 with CMEs, 
reporting that 97\% of them were associated with fast Earth-directed full-halo events. 
More extensive statistical analyses addressing the relation between SEPs and CMEs were conducted by 
\citet{dierckxsens:15} ($> 160$ SEP events in Solar Cycle 23), 
\citet{papaioannou:16} (314 SEP events from 1984 to 2013), 
and \citet{paassilta:17} (176 SEP events in Solar Cycles 23 and 24), 
all evidencing the role of halo CMEs as major SEP generators.
Considering the SEP generation as consequence of CME interactions \citep{gopal:02},
the most favourable condition for SEP production appears to be the case of a preceding, relatively slow CME which is caught up by a second, faster CME launched some hours later. The optimal time interval between two subsequent CME eruptions, in order to have their intersection close to the Sun ($\le 20 \,\, R_\odot$), has been found to be $\ge 7$ hours, so to have a maximum acceleration efficiency at $\sim 5-15 \,\, R_\odot$.
This so-called “twin-CME” scenario, however, appears controversial and
it has been debated by \citet{kahler:14}, who questioned the role of CME interactions in major SEP events
finding that they can be explained by a general increase of both background seed particles and more frequent CMEs during times of higher solar activity.
Overall, previous results have confirmed the major role of fast Earth-directed halo CMEs and interacting CMEs as sources of strong SEP events at Earth,
and represent an additional reason to study these types of CMEs as the most potentially geoeffective.

For this reason, we consider of primary importance to further study the relationship between CMEs, 
their interplanetary counterparts, and the triggered geomagnetic activity in the latest years as well.
In the past years, several works have tried to construct storm prediction models using
empirical relations linking real-time in-situ solar wind parameters to geomagnetic activity indices (\cite{obrien:00b}; \cite{newell:07}, \cite{newell:08})
or statistical analyses of in-situ ICME properties and geomagnetic storms (\cite{srivastava:04}; \cite{zhang:07}).
Although near-Earth solar wind parameters can yield a quite reliable prediction of geomagnetic storm events, for spacecraft located at the Lagrangian point L1 such kind of warnings can give only a one-hour notice.  
To overcome this forecasting limit, studies based on remote observations of CME parameters and their association with other kinds of solar activity features have also been carried out, proving to be a powerful complementary approach to solar wind-magnetospheric coupling functions and statistical ICME studies (\cite{dumbovic:15}).
A third approach, complementary to the other two, is the use of global 3D MHD simulations to model the CME propagation in the heliosphere from Sun to Earth,
thus also providing a determinant support in studies of CME propagation and multiple CME interactions.
In this sense, combining the three approaches mentioned before into a single methodological scheme analysis may lead to a more comprehensive space weather forecasting tool to be used not only at Earth location but possibly also at other spacecraft and planetary locations.

\medskip
In this study we address the problem of predicting the geoeffectiveness of halo CMEs by reconstructing their Sun-to-Earth global evolution by means of 3D global MHD simulations, using then the simulation outputs upstream of Earth to predict their impact on geospace.
Moreover, from the analysis of their solar source regions we look for recurrent features that can be used as precursors of CME-driven geomagnetic storms.
We focus on Solar Cycle 24 because of the relatively limited number of statistical studies available
compared to Solar Cycle 23 (see for example \cite{richardson:13}; \cite{watari:17}), 
as well as because of its peculiar characteristics - weak solar activity and mild space weather \parencite{gopal:15a}.

We first select a set of fast halo CMEs observed by the SOHO/LASCO instrument over an 81-month period during Solar Cycle 24 (Jan. 2009-Sep. 2015).
We use multi-instrument observations of the solar photosphere and low corona to investigate their association with other solar activity features.
In the attempt of reconstructing the propagation and global evolution of each event up to its arrival at Lagrange point L1, we made use of the WSA-ENLIL+Cone model \parencite{odstrcil:03} running at NASA/CCMC (\url{https://ccmc.gsfc.nasa.gov}), which represents the currently most widely used code for CME/ICME modelling; 
simulation outputs at Earth are then compared with in-situ measurements provided by the Wind spacecraft.
Finally, using the coupling function proposed by \citet{newell:08} to link upstream solar wind conditions to the global $K_p$ index, we estimate the geomagnetic activity level and compare it with global data records.

In Section \ref{sec:event_selection} we describe the CME selection procedure from the SOHO/LASCO halo CME catalogue and the complementary data used.
CME modelling with WSA-ENLIL+Cone is described in Section \ref{sec:cme_modelling}.
In Section \ref{sec:prediction_geoeffectiveness} we present our geoeffectiveness prediction scheme.
Results of the statistical analysis of the selected event properties, evolution, associated events and impact on geospace are then discussed in Section \ref{sec:results_solar} (solar source regions), \ref{sec:results_ip} (interplanetary signatures and shock association), \ref{sec:results_geo} (impact on geospace) and \ref{sec:sep} (association with major SEP events). Conclusions are presented in Section \ref{sec:conclusions}.

\section{Event selection and complementary data}
\label{sec:event_selection}

Earth-directed halo CMEs observed from coronagraphs along the Sun-Earth line are the most geoeffective CME type. 
For this reason, the SOHO/LASCO halo CME catalogue (\url{http://cdaw.gsfc.nasa.gov/CME_list/halo/halo.html}, \citealt{gopal:10}) has been the primary database used in this work to identify and select the CME events of interest. 

To current knowledge, the most important parameters assessing the potential geoeffectiveness of a CME from coronagraphic observations are their source location and reconstructed speed in 3D space (\cite{michalek:06}; \cite{cid:12}; \cite{dumbovic:15}). 
For this reason we have estimated the CME speed in 3D space $V_\textup{space}$ starting from coronagraphic observations of the speed projected on the sky plane $V_\textup{sky}$ by considering the early CME evolution as characterised by a fixed angular width (\emph{e.g.} as in a cone model). 
To link $V_\textup{sky}$ to $V_\textup{space}$ we applied the relation proposed by \citet{gopal:10}
\begin{equation}
\centering
V_\textup{space} = \frac{\cos{\omega} + \sin{\omega}}{\cos{\omega} \cos{\Theta} + \sin{\omega}} \, V_\textup{sky},
\label{eqn:v_space}
\end{equation}
where $\Theta$ is the angle between the cone axis and the sky plane and $\omega$ is the CME half-width angle.\\
The CME source location was derived from the Solar and Geophysical Activity Summary (SGAS) listing (available at \url{ftp://ftp.swpc.noaa.gov/pub/warehouse/}), defined as the heliographic coordinates of the associated {H}$_\alpha$ flare.
If no such data were available, source information was obtained from solar disk images such as Yohkoh/SXT, SOHO/EIT or from other data such as microwave images from the Nobeyama radioheliograph and {H}$_\alpha$ images.
As an evaluation of the half width for halo CME events is not possible from single-spacecraft observations, and average half-width angle of the CME was derived from an empirical relation proposed by \citet{gopal:09b}, who reported a correlation of $0.69$ between the sky plane speed and the angular width $w$ of a CME
\begin{equation}
V_\textup{sky} = 360 + 3.62 \cdot w \,\, [\si{\kms}],
\end{equation}
for a set of 341 near-limb CMEs (for which the evaluation of the width is easier). 
Even though for halo CMEs a direct estimate of the width is extremely difficult, starting from the previous set of limb events an average half-width angle $\omega$ has been identified. Adapting this correlation to halo CMEs,
$\omega=32^\circ$ for $V_\textup{space}\le 500$ \si{\kms},
$\omega=45^\circ$ for $500$ \si{\kms} $< V_\textup{space} \le 900$ \si{\kms} and 
$\omega=66^\circ$ for $V_\textup{space} > 900$ \si{\kms}. 
The errors associated to the various CME parameters have been estimated to be $dV_{sky}/V_{sky} = 10$\%, $d\omega = 10^{\circ}$ and $d\Theta = 10^{\circ}$ (\cite{gopal:10}; \cite{jang:16}). 

\medskip
Limiting our interest to the period Jan. 2009-Sep. 2015, we have selected all the fast ($V_\textup{space} \ge 1000$ \si{\kms}) halo CMEs that originated from a heliographic {longitude} between $30^{\circ}$ E and $30^{\circ}$ W from the solar central meridian.
The application of these criteria to halo CMEs listed in the catalogue resulted in a set of 21 events hereafter denoted as ``S1''.
As fast CMEs can be deflected towards the east due to the blocking effect of the solar wind background ahead of them \citep{wang:04},
we have extended the source location condition towards the western hemisphere so to account for potentially geoeffective CMEs undergoing such kind of deflection. 
In particular, we have selected all the fast halo CMEs that originated from a heliographic {longitude} between $30^{\circ}$ W and $60^{\circ}$ W from the solar central meridian, resulting in a set of 9 events denoted as ``S2''.
To be sure of including in our analysis all the geoeffective halo CME events that occurred during Solar Cycle 24, 
we have checked the $K_p$ index data for the whole period considered (available at \url{http://www.gfz-potsdam.de/en/kp-index/}), 
and we have investigated the active periods characterised by a 3-hour $K_p \ge 5$. 
For all the geomagnetic activity periods above the threshold, assuming a CME propagation time from Sun to Earth of $\sim$2-3 days, we have searched the catalogue for CME precursors not included in the S1 and S2 sets, which nevertheless resulted in a strong geomagnetic storm. 
In addition, we have also selected all the halo CMEs that originated from the same solar region and within 2 days before and after the events already included in the sets S1 and S2 (imposing no constraint on the CME speed in this case). This choice was made in order to take into account CME-CME interactions and pre-conditioning of the solar wind background due to the passage of previous CMEs, factors that may also affect the propagation and geoeffectiveness of the single events involved (\citealt{burlaga:02}; \citealt{temmer:17}).
The application of these two criteria has resulted in a set of 23 additional CMEs named ``S3'', which includes:
(a) all halo CMEs originated from the east ($90^{\circ}$ E $<$ longitude $<$ $30^{\circ}$ E) and the west ($60^{\circ}$ W $<$ longitude $<$ $90^{\circ}$ W) part of the visible disk that resulted in a geomagnetic storm ($K_p \ge 5$);
(b) halo CMEs that took part in interactions/pre-conditioning of the solar wind background - in particular those recorded within $\pm 2$ days from the halo events in S2 and S3.
It is worth noticing that all the S3 events satisfy one of the two geoeffectiveness conditions (source location or speed condition).

A total number of 53 CMEs composed the final set of events that have been analysed in this work.
Table~\ref{tab:events} presents a complete list of the selected events, together with their main observational properties as reported in the LASCO halo CME catalogue.
Figure~\ref{fig:source_location} shows the distribution of the source locations of the selected CMEs on the solar disk, in heliographic coordinates.
The distribution appears symmetric with respect to the solar equator with the majority of the events (85\%) originating within a latitude of $\pm$($10^{\circ}-30^{\circ}$).
Moreover, none of the selected events originated from a latitude higher than $\pm \, 35^{\circ}$.
%

\begin{figure}[t]
\centering
   {\includegraphics[width=.60\textwidth]{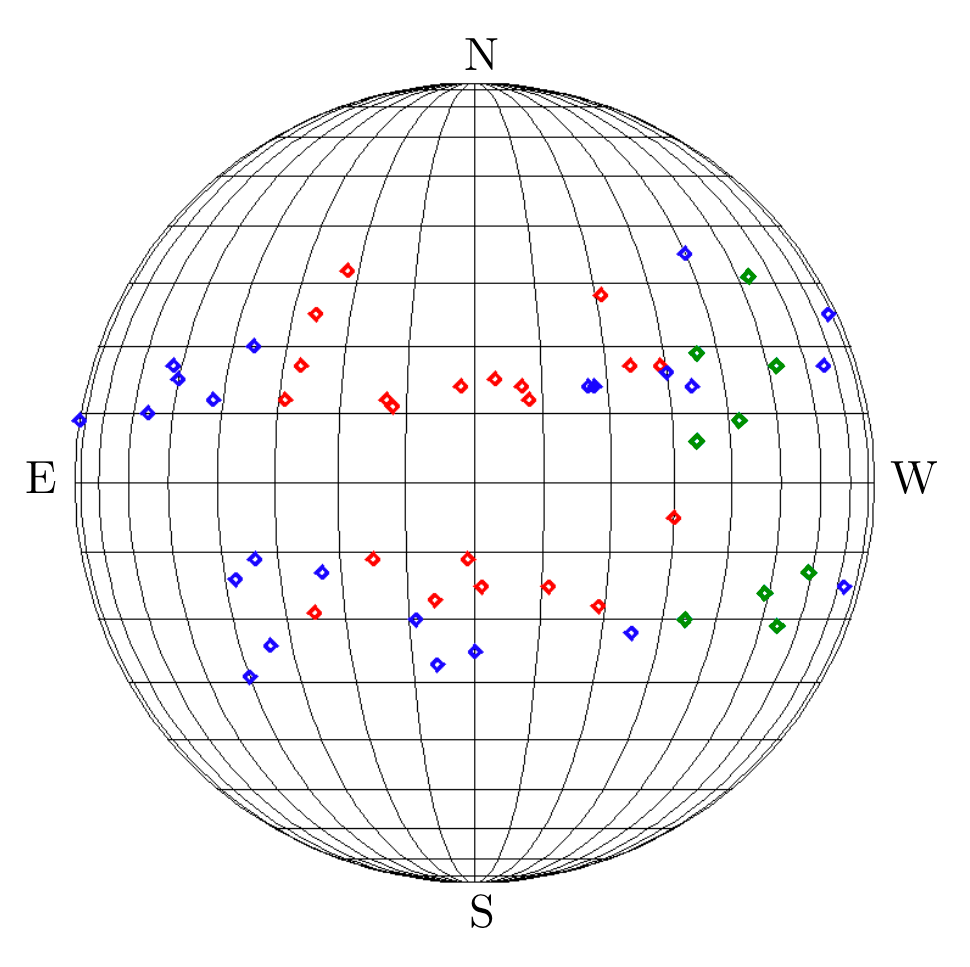}} 
\caption[Source locations of selected CMEs in heliographic coordinates]{Distribution of the source locations of the selected CME events in heliographic coordinates. Red, green and blue diamonds indicate CMEs belonging to S1, S2 and S3 respectively.}
\label{fig:source_location}
\end{figure}

\newpage
	\begin{center} 
	\begin{longtable}{cccccccc}
	\toprule
	\multicolumn{8}{c}{\textbf{Selected CMEs}} \\
	\midrule
	Event 			& Set	& Start date			& Start time 	& Space speed 	&\multicolumn{2}{c}{Source location} 	& Associated flare	\\
	 \#				&		& (YYYY-MM-DD)	& (UT)		& (\si{\kms})		&\multicolumn{2}{c}{(HEEQ)}			& (SXR class)      \\
	\midrule
	1				&	3	& 2010-04-03		& 10:33 		& 939			&-25  	&	0					&B7.4			\\
	2				& 	3	& 2010-08-01		& 13:42 		& 1030			&20	&	-36					&C3.2			\\
	3				& 	2	& 2010-08-14		& 10:12			& 1280			&17	 	&	52					&C4.4		\\
	4				&  	2	& 2011-03-07		& 20:00		& 2223			&31 	&	53					&M3.7				\\
	5				&  	1	& 2011-06-02		& 08:12			& 1147			&-19 	&    -25					&C3.7			\\
	6				&  	2	& 2011-06-07		& 06:49		& 1321			&-21 	&	54					&M2.5				\\	
	7				&  	3	& 2011-08-03		& 14:00			& 785			&16	 	&	30					&M6.0				\\	
	8				&  	2	& 2011-08-04		& 04:12			& 1477			&19	 	&	36					&M9.3				\\	
	9				&  	1	& 2011-09-06		& 02:24		& 1232			&14 	&	7					&M5.3				\\	
	
	10				&  	3	& 2011-09-06		 & 23:05		& 830			&14	 	&	18					&X2.1				\\
	11				&  	3	& 2011-09-22		&10:48			& 1905			&9		&	-89					&X1.4				\\	
	12				&  	3	& 2011-09-24		& 12:48			& 2018			&10 	&	-56					&M7.1			\\	
	13				&  	3	& 2011-09-24		& 19:36			& 1076			&12	 	&	-42					&M3.0				\\	
	14				&  	3	& 2011-10-22		& 01:25			& 666			&35	&	40					&-				\\	
	15				&  	3	& 2011-01-22		& 10:24			& 1011			&25	 &	77					&M1.3				\\	
	16				&  	1	& 2012-01-19 		& 14:36			& 1269			&32	 &	-22					&M3.2				\\	
	17				&  	1	& 2012-01-23		& 04:00		& 2511			&28	 &	21					&M8.7			\\	
	18				&  	3	& 2012-03-05		&04:00			& 1627			&17	 	&	-52					&X1.1				\\	
	19				&  	1	& 2012-03-07		& 00:24		& 3146			&17	 	&	-27					&X5.4				\\	
	
	20				&  	1	& 2012-03-07		&01:30			& 2160			&25	 &	-26					&X1.3				\\	
	21				&  	1	& 2012-03-09		& 04:26		& 1229			&15	 	&	3					&M6.3				\\	
	22				&  	1	& 2012-03-10		&18:00			& 1638			&17	 	&	24					&M8.4				\\		
	23				&  	3	& 2012-03-13		& 17:36			& 1931			&17	 	&	66					&M7.9				\\	
	24				&  	3	& 2012-04-23		& 18:24			& 769			&14	 	&	17					&C2.0			\\		
	25				&  	1	& 2012-06-14		& 14:12			& 1254			&-17	&	-6					&M1.9				\\	
	26				&  	3	& 2012-07-04		& 17:24			& 830			&14 	&	34					&M1.8				\\	
	27				&  	2	& 2012-07-06		& 23:24		& 1907			&-13	&	59					&X1.1				\\
	28				&  	1	& 2012-07-12		& 16:48			& 1405			&-15	&	1					&X1.4				\\	
	29				&  	2	& 2012-09-28		& 00:12			& 1093			&6		&    34					&C3.7				\\
		
	30				&  	1	& 2013-03-15		& 07:12			& 1366			&11	 	&	-12					&M1.1			\\	
	31				&  	1	& 2013-06-28		& 02:00		& 1254			&-18	&    19					&C4.4				\\	
	32				&  	1	& 2013-08-17		& 19:12			& 1418			&-5 	&	30						&M1.4				\\	
	33				&  	1	& 2013-09-29		& 22:12			& 1370			&17	  	&	29					&C1.3				\\	
	34				&  	3	& 2013-11-07		& 15:12			& 626			&-13	&   -23					&M2.4				\\	
	35				&  	2	& 2013-12-07		&07:36			& 1165			&-16	&    49					&M1.2				\\	
	36				&  	3	& 2014-01-04		& 21:22			& 1166			&-11	&	-34					&M4.0			\\	
	37				&  	1	& 2014-01-07		&18:24			& 2246			&-15	&	11					&X1.2				\\	
	38				&  	1	& 2014-02-16		& 10:00			& 1064			&-11	&	-1					&M1.1			\\	
	39				&  	3	& 2014-02-18		& 01:36			& 942			&-24	&	-34					&-				\\	
	
	40				&  	3	&	2014-02-20		& 03:12		& 1115			&-14	&	-38					&C3.3			\\	
	41  				&  	3	&	2014-02-20		& 08:00	& 960			&-15	&	73					&M3.0				\\	
	42				&  	2	&	2014-04-18		& 13:25		& 1359			&-20	&	34					&M7.3				\\
	43				&  	3	&	2014-06-04		& 12:48		& 555			&-29	&	-40					&-			\\
	44				&  	1	&	2014-09-09		& 00:06	& 1080			&12		&    -29					&M4.5				\\
	45				&  	1	&	2014-09-10		& 18:00		& 1652			&14		&	-2					&X1.6			\\	
	46				&  	3	&	2014-12-17			&05:00		& 855			&-20	&	-9					&M8.7				\\	
	47				&  	1	&	2014-12-19			& 01:04		& 1513			&-11	&	-15					&M6.9			\\	
	48				&  	3	&	2015-03-15		& 01:48		& 932			&-22	&	25					&C9.1			\\	
	49				&  	3	&	2015-06-18		&17:24		& 1398			&15		&    -50					&M3.0				\\	
	
	50				&  	3	&	2015-06-19		& 06:42	& 798			&-27	&	-6					&-			\\
	51				&  	1	&	2015-06-21		& 02:36	& 1740			&12		&    -13					&M2.0				\\	
	52				&  	1	&	2015-06-22		& 18:36		& 1573			&12		&    8					&M6.5				\\	
	53				&  	2	&	2015-06-25		& 08:36	& 1805			&9		&    42					&M7.9				\\	
	\bottomrule
	\\
	\caption[Complete list of the selected CME events]{Complete list of the selected CME events. Columns 3 and 4 refer to the first appearance in LASCO C2 coronagraph. Column 5 lists the speed in 3D space calculated using Equation \ref{eqn:v_space}. Columns 6 and 7 report the heliographic latitude and longitude of the source location. Column 8 reports the SXR class of the associated flares, when available; ``-'' indicates no associated flare or an association with a weak B- or A-class event.}
	\label{tab:events}
	\end{longtable}
	\end{center} 

\subsection{Reconstructing the global scenario by means of remote-sensing and in-situ data} 
To fully reconstruct the Sun-to-Earth evolution of the selected CME events, we have made use of complementary data archives containing both remote sensing observations of the Sun and in-situ measurements of the solar wind plasma and magnetic properties.

Searching for potential precursors of upcoming geoeffective CMEs, we have checked the association of all the selected CMEs with additional solar activity features observed by means of remote-sensing instruments. 
To investigate the properties of the CME solar source regions and their association with solar flares, we have made use of the data provided by NOAA/SWPC
and listed in the Solar Region Summary (SRS)
and in the Solar and Geophysical Activity Summary (SGAS) (\url{ftp://ftp.swpc.noaa.gov/pub/warehouse/}).

With the aim of investigating pre- and post-eruptive conditions at the Sun, 
we have searched for associations with filaments/prominences, X-ray sigmoidal structures and global coronal perturbations.
In performing this association check we have made use of the iSolSearch interactive tool provided by Heliophysics Events Knowledgebase (HEK), a database collecting data acquired primarily by SDO/AIA and SDO/HMI instruments (\cite{lemen:12}; \cite{scherrer:12}) and available at \url{http://www.lmsal.com/isolsearch}.
In determining whether an association exists, we imposed both temporal and spatial criteria,
namely that a given activity feature was observed in the same NOAA AR of the reconstructed CME source location, 
and that it occurred within $\pm 60$ minutes from the CME onset time. 
The onset time of each CME event was estimated by back-extrapolating to the solar surface the height-time information contained in the LASCO catalogue, using a linear fitting.

To monitor the solar wind conditions right before the impact on the magnetosphere, we have made use of in-situ data obtained by the Wind spacecraft via two of its on-board instruments: 
the Magnetic Field Investigation (MFI) 
and the Solar Wind Experiment (SWE) (\cite{lepping:95}; \cite{ogilvie:95}).
MFI and SWE solar wind data products and derived quantities were used as comparison with the ENLIL simulation results at Earth as discussed in Sections~\ref{sec:cme_modelling}. In particular, in the comparison phase we used 1-minute resolution data relative to the solar wind magnetic field strength $B$, solar wind (proton) bulk speed $V$, proton number density $N$, proton temperature $T$, and $\beta$ factor. 

In order to further characterise the effects of the selected CME events on Earth, we have analysed their association with major SEP events listed in the NASA (\url{http://cdaw.gsfc.nasa.gov/CME_list/sepe/}, containing data up to end 2014) and NOAA (\url{http://umbra.nascom.nasa.gov/SEP/}) catalogues
\citep{reames:99}. 
Major SEP events are those associated with intensities $\ge 10$ pfu\footnote{pfu = particles $\cdot$ \si{\cm^{-2}}$\cdot$ \si{\s^{-1}}$\cdot$ sr$^{-1}$.} in the $\ge 10$ \si{\mega \eV} proton energy channel, and are the most relevant events in terms of space weather effects (\cite{schwenn:05}; \cite{gopal:15c}).

\section{CME modelling with WSA-ENLIL+Cone}
\label{sec:cme_modelling}

The ENLIL model is a global 3D ideal MHD code that models the evolution of the background solar wind plasma and magnetic field in the heliosphere up to $10$ AU (\cite{odstrcil:03}; \cite{toth:odstrcil:96}). 
In this work we have used ENLIL version 2.8f, currently running at NASA/CCMC and available for runs on request,
in combination with the Wang-Sheeley-Arge ({WSA}) empirical coronal model \parencite{arge:00},
which takes as input synoptic magnetograms (Carrington maps) from the National Solar Observatory (NSO) and the Global Oscillation Network Group (GONG).

The WSA coronal model is made by the combination of the WSA Potential Field + Current Sheet (WSA PF+CS) and the WSA Inner Heliosphere (WSA-IH) models. 
The WSA PF+CS model combines a Potential Field Source Surface (PFSS) model with the Schatten Current Sheet model to model the magnetic field between the photosphere and the source surface, set at $2.5$ $R_\odot$  \parencite{owens:13}. 
Starting from $2.5$ $R_\odot$ outwards, the WSA-IH model then propagates the solar wind and magnetic field up to $21.5$ $R_\odot$, where they are used as the inner boundary conditions for the heliospheric model.
For the heliospheric model, we used a simulation domain between $21.5$ $R_\odot$ and $2$ AU in the radial direction, so to include the Earth orbit,
with a latitudinal angle ($\theta$) going from $60^{\circ}$ N to $60^{\circ}$ S with respect to the solar equator, and an azimuthal angle ($\phi$) spacing over $360^{\circ}$.

\medskip
Run in combination with a cone model (\cite{zhao:02}; \cite{xie:04}), ENLIL can model the propagation of CMEs throughout the heliosphere.
This kind of simplified model assumes a self-similar CME expansion, characterised by a constant angular width as result of the external magnetic pressure confinement, until the CME reaches the heliospheric inner boundary at $21.5 \, R_\odot = 0.1$ AU - as supported by coronagraphic observations \citep{stcyr:00}.
 
The WSA-ENLIL+Cone model takes as CME input parameters: the passage time at $21.5 \, R_\odot$, the radial speed at $21.5 \, R_\odot$, its direction of propagation, and its half width. 
The date and time of the CME passage at the ENLIL inner boundary have been obtained 
by de-projecting the linear $V_\textup{sky}$ contained in the LASCO catalogue by means of Equation~\ref{eqn:v_space}, and assuming that $V_\textup{space}$ was maintained constant from the CME onset to the boundary. 
Under this assumption, we have extrapolated the passage time at $21.5$ $R_\odot$ starting from the estimated onset time contained 
in the height-time plots in the LASCO halo CME catalogue, assuming that the ENLIL cone originated at a distance of about $1$ $R_\odot$ from the solar centre (\emph{e.g.} near the solar surface or in the chromosphere). The obtained $V_\textup{space}$ has been used as input for the radial velocity parameter, assuming a completely radial direction of the CME cones at the inner boundary.

Assuming a radial propagation in the corona, the CME direction of propagation has been specified by the latitude and longitude of the CME source location as listed in the LASCO halo CME catalogue.
The half-width angle has been estimated by means of the empirical relation proposed by \citet{gopal:09b} and presented in Section \ref{sec:event_selection}. 
Note that the ENLIL+Cone model version used in this work did not take into account the CME internal magnetic structure, so that the CME blobs can only perturb the pre-existing interplanetary magnetic field background once they are inserted in the heliospheric domain. 

In addition to these major input parameters, the structure of a cone CME can also be adjusted by the user 
by specifying the shape of the cone base and that of the CME cloud on top of it.
For the sake of simplicity and to avoid the introduction of too many parameters in the simulation runs, in this work a spherical shape has been used in all runs.

Additional parameters that can be set by the user involve the solar wind background properties. In particular, the density enhancement factor (``df factor'') of the CME with respect to the solar wind background has been initially set equal to 4 (default value) for all the events. However, being aware that \citet{taktaki:09} reported a better performance in the case of df $=2$ when validating the model, the March 2012 (\#18-23) and June 2015 (CMEs \#49-53) sequences have been simulated under both conditions. 
For these two events, the choice df $=2$ has performed better in terms of arrival times and peak values at Earth, and therefore in the following discussion we have considered such runs only. Although further testing would be needed to confirm this point, we report that this result seems to affect in particular CMEs propagating through a pre-conditioned background, \emph{e.g.} the case of CME sequences. 
To our knowledge, no publication exists investigating the effect of such parameter in the case of complex CME events, and although a better performance has been reported by \citet{taktaki:09} in the case of single CMEs, no conclusive explanation of this effect in simulations has been reported so far.

\section{Geoeffectiveness prediction scheme and forecast verification}
\label{sec:prediction_geoeffectiveness}

To predict the geoeffectiveness of CME events starting from solar observations and heliospheric simulations,
it is crucial to link ENLIL simulation outputs to the expected level of perturbation induced on geospace.
To do this, we have assumed an operational definition of geoeffectiveness using both (1) the solar wind-geomagnetic activity coupling function approach and (2) the classical method based on the evaluation of the dayside magnetopause stand-off distance in the Sun-Earth direction.
The chart reported in Figure~\ref{fig:scheme} illustrates the prediction scheme discussed below. 

\medskip
\emph{Geomagnetic activity indices.}
An approach to estimate the geoeffectiveness of a CME is by means of empirical relations linking the solar wind plasma and magnetic field parameters upstream of Earth to its magnetospheric effects, considering for instance the geomagnetic activity indices most widely used.
%
\citet{newell:07} showed that all the most commonly used geomagnetic indices, including the $K_p$ index, 
can be related to solar wind parameters by the use of the following coupling function, estimating the rate of magnetic field lines opened at the magnetopause
\begin{equation}
\label{eqn:reconnection_rate}
\frac{d \Phi_\textup{MP}}{dt}= V^{4/3} \, B^{2/3} \, \sin^{8/3}({\theta_{c} /2}),
\end{equation}
where 
$\theta_c = \arctan{(B_y/B_z)}$ is called the \emph{IMF clock angle} and indicates the direction of the IMF: 
$B_z$ refers to the north-south component of the IMF relative to Earth and 
$B_y$ refers to the component of the IMF perpendicular to both the Sun-Earth line and the north-south line in geocentric solar magnetospheric (GSM) coordinates.
$\theta_{c}=0$ corresponds to a completely northward oriented magnetic field, while $\theta_{c}=\pi$ corresponds to a completely southward oriented magnetic field. \\
The $K_{p}$ index can be predicted once the speed and the magnetic properties of the solar wind are known, by means of the following relation (with an $r$ factor $=0.866$)
\begin{equation}
\label{eqn:kp}
K_{p} = 0.05 + 2.224 \cdot 10^{-4} \,  \frac{d \Phi_\textup{MP}}{dt}  + 2.844 \cdot 10^{-6} {N}^{1/2} V^{2},
\end{equation}
where $V$ is measured in \si{\kms}, $N$ is the solar wind number density and is measured in \si{ \cm^{-3}} and $B$ (included in the reconnection rate term) is in \si{\nano\tesla} \citep{newell:08}. 
As the ENLIL+Cone model used in this work does not take into account the internal magnetic structure of CMEs/ICMEs, 
it is incapable of reproducing the magnetic features generally observed in association with ICMEs \emph{e.g.} turbulent sheaths and magnetic clouds.
For this reason, we have decided not to consider simulation outputs for the $\bm{B}$ field orientation as reliable enough to be used in our prediction scheme.
To calculate the predicted $K_p$ index values we have made use of Equation~\ref{eqn:kp} 
under the assumption of two orientation conditions for the ICME structure impacting on Earth:
\begin{itemize}
\item[1)] Completely southward magnetic field, consistent with a worst-case scenario in terms of CME geoeffectiveness (maximum impact on geospace): 
$\theta_c = \pi$ and therefore $\sin^{8/3}(\theta_c/2) = 1$.
\item[2)] Randomly oriented magnetic field: assuming a uniform distribution of the clock angle $\theta_c $, the $\sin^{8/3}(\theta_c/2)$ term entering Equation~\ref{eqn:reconnection_rate} has an expectation value $\langle \sin^{8/3}(\theta_c/2) \rangle \sim 0.45$ \parencite{emmons:13}. 
\end{itemize}
%

\bigskip
\emph{Magnetopause stand-off distance.}
Another important parameter assessing the geoeffectiveness of a CME impacting on Earth 
is the magnetospheric compression due to the dynamic pressure exerted by the solar wind on the magnetosphere.
A first-order evaluation of this compression is given by the {magnetopause stand-off distance}, defined as the position of the magnetopause nose along the Sun-Earth direction.
Assuming a dipolar shape of the magnetosphere on the dayside, a rough estimate of the magnetopause stand-off distance is given by \parencite{taktaki:09}
	\begin{equation}
	\label{eqn:magnetopause_standoff}
	d_{\textup{so}}= \left (    \frac{B_0^2}{2 \cdot 0.88 \, \mu_0 \,  \rho \,  V^2}   \right)^{1/6} \cdot R_\textup{E}.
	\end{equation}
In this case ENLIL outputs can be directly used to calculate the expected magnetopause stand-off distance.

\begin{figure}[t]
\centering
\resizebox{.99\linewidth}{!}{
\begin{tikzpicture}[node distance = 1.4cm, shorten >=.5 pt]

\scriptsize

\node [model, text width=9em] (model1) {SOHO/LASCO CME observations};
\node [block, below of=model1, node distance = 1cm, text width=9em] (A2) {WSA-ENLIL+Cone};

\node [block, below of=A2, node distance = 1cm, text width=3em] (B4) {${V}$};
\node [block, left of=B4, node distance = 2cm, text width=3em] (B3) {${B}$};
\node [block, right of=B4, node distance = 2cm, text width=3em] (B5) {$\rho$, $N$};

\node [output, below of=A2, node distance = 3.5cm, text width=12em] (C1) {Predicted magnetopause stand-off distance};
\node [output, right of=C1, node distance = 4cm] (C2) {Predicted $K_p$ index};
\node [block, text width=14em, above of=C2, node distance = 1cm] (A4) {Coupling functions \\
																	solar wind-geomagnetic indices};

\draw [->] (model1) -- (A2);
\draw [->] (A2) -- (B3);
\draw [->] (A2) -- (B4);
\draw [->] (A2) -- (B5);

\draw [->,shorten >=3pt] (B3) -- (A4);
\draw [->,shorten >=5pt] (B4) -- (A4);
\draw [->,shorten >=5pt] (B5) -- (A4);
\draw [->] (B3) -- (C1);
\draw [->] (B4) -- (C1);
\draw [->] (B5) -- (C1);
\draw [->] (A4) -- (C2);

\end{tikzpicture}}
\caption[Geoeffectiveness prediction scheme applied to the selected CME events]{Geoeffectiveness prediction scheme used to evaluate the expected impact at Earth of the selected CME events.}
\label{fig:scheme}
\end{figure}
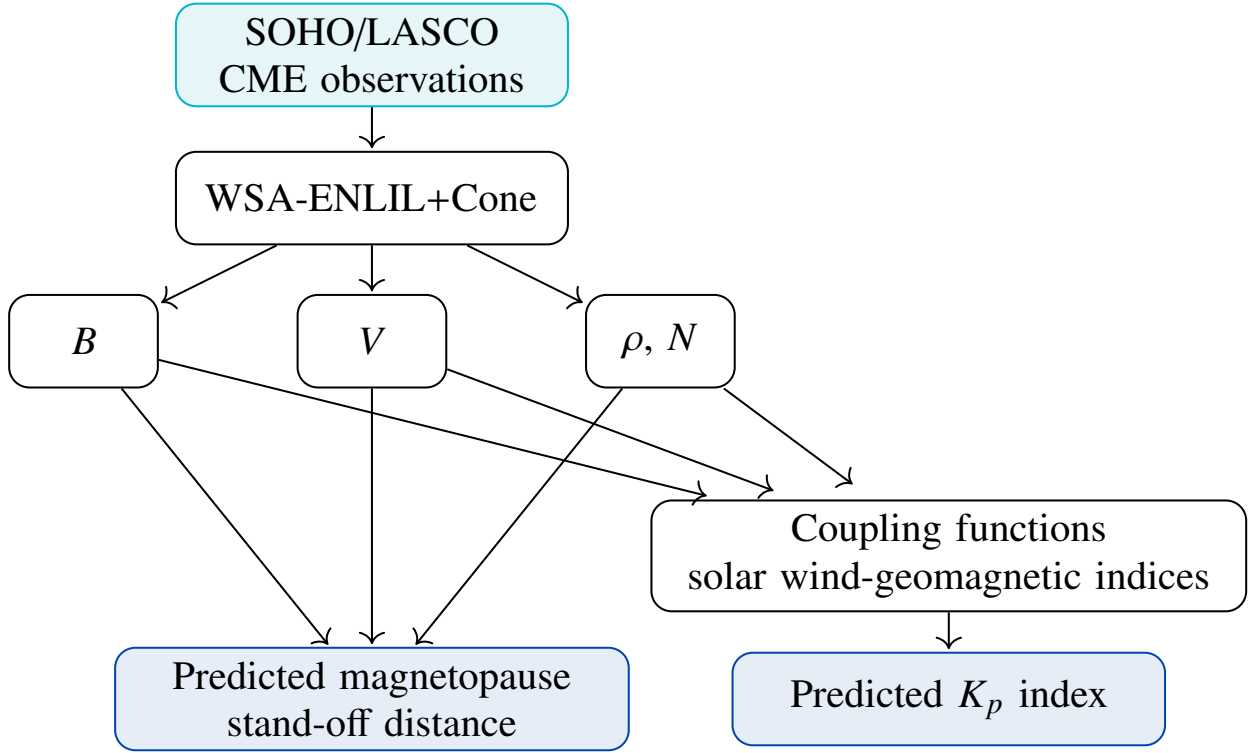

\medskip
\emph{Forecast verification.}
To quantitatively evaluate the performances of the geoeffectiveness prediction scheme discussed above, 
we have computed the 2x2 contingency tables for categorical forecasts (\cite{weigel:06, jolliffe:steph:11})
relative to two main observables: the CME arrival at Earth location and the CME induced perturbation on the geomagnetic field.  
Categorical forecasts are those associated with two possible outcomes: yes (event) and no (no event). 
Starting from the results of our prediction scheme, 
in this work we have considered the following two ``yes'' forecast conditions as independent conditions for CME geoeffectiveness:
(a) CME arrival, marked in-situ by the detection of a ICME-driven forward shock;
note that as shocks can extend much further than ICMEs, cases where the ICME missed the Earth but the forward shock was detected in-situ were considered as ``yes'' events (see previous works on ``driverless shocks'' by \citet{gopal:10b} and \citet{janvier:14}).
Note that in this work we have focused on the detection of IP shocks as signatures of ICME arrival as this was the only way to compare ENLIL results to in-situ measurements. 
In fact, as ENLIL+Cone treats CMEs as hydrodynamics plasma blobs inserted in the solar wind background field, it does not account for the CME internal magnetic structure and hence it is not able to provide information about the ICME/MC passage following a shock at Earth.
(b) Geomagnetic storm condition defined as characterised by a $K_p \ge 5$ consistent with NOAA Space Weather Scale for geomagnetic storms.
The contingency table for each forecast is characterised by the following entries, 
calculated by comparing forecasts to observations: 
\emph{hits}, defined as events that were both predicted and observed to occur; 
\emph{misses}, defined as events that were not predicted, but were observed to occur; 
\emph{false alarms}, defined as events that were predicted to occur, but were observed not to occur; 
and \emph{correct negatives}, events that were correctly predicted not to occur.
Starting from the entries in the contingency tables for each observable, in order to provide a quantitative evaluation of the forecast performances we computed: 
(a) the correct rejection rate (CR rate), defined as the number of correct negatives divided by the number of observed ``no'' events 
and addressing the question \emph{``what fraction of the observed ``no'' events was correctly predicted not to occur?''};
(b) the false alarm rate (FA rate), defined as the number of false alarms divided by the number of observed ``no'' events
and addressing the question \emph{``what fraction of the observed ``no'' events was incorrectly predicted to occur?''};
(c) the correct alarm ratio (CA ratio), defined as the number of hits divided by the number of predicted ``yes'' events
and addressing the question \emph{``what fraction of the predicted ``yes'' events did actually occur?''};
and (d) the false alarm ratio (FA ratio), defined as the number of false alarms divided by the number of predicted ``yes'' events
and addressing the question \emph{``what fraction of the predicted ``yes'' events actually did not occur?''}.

\section{Analysis of solar conditions}
\label{sec:results_solar}

\subsection{Association with active regions}

Over a set of 53 CMEs under study, 47 of them ($89$\%) originated from an active region. 
Previous studies on slower CME samples, reported lower CME-AR association rates: 
on the one hand, a $63$\% association was found by \citet{chen:11} over a set of 224 CMEs observed by the LASCO instrument in the years 1997-1998, 
using data relative to 108 ARs obtained by the SOHO Michelson Doppler Imager (MDI);
on the other hand, \citet{subramanian:01} found an $85$\% association studying a set of 32 front-side halo (full and partial) CMEs observed by LASCO over the same period, primarily using AR data from SOHO/EIT and SOHO/MDI. 

\medskip
\emph{Mount Wilson classification.} 
Considering the Mount Wilson classification of ARs, the most common classes were $\beta\gamma\delta$ and $\beta\gamma$ (18 and 15 events respectively), followed by simple $\beta$ topologies (8 events) and simple $\alpha$ topologies (5 events). There was only one AR showing a $\beta\delta$ topology and none representing other classes. 
No information on the AR associated with CME \#27 was available, so this event has not been considered in the following discussion.
As shown in Figure~\ref{fig:4_mountwilson}, comparing our distribution with the base distribution reported by \citet{jaeggli:16} we found a significant enhancement in the fraction of $\beta\gamma\delta$ and $\beta\gamma$ regions observed, while $\alpha$ and $\beta$ classes resulted significantly depressed in their occurrence frequency. 
The fraction of $\beta\delta$ regions remained almost unchanged.

These results show that $\beta\gamma\delta$ and $\beta\delta$ configurations are characterised by the highest rate of production for fast CMEs, 
confirming the interpretation that AR with complicated magnetic configurations have a higher probability to undergo an energetic eruption that releases the magnetic energy stored over a short time period.
Such kind of analysis has been previously conducted on solar flares by \citet{qu:08}, who reported a similar result in terms of flare production rates.
However, to our knowledge, no previous literature exists in the case of CMEs, not for Cycle 24 nor for previous ones.

    \begin{figure}
   \centering
   {\includegraphics[width=0.80\columnwidth]{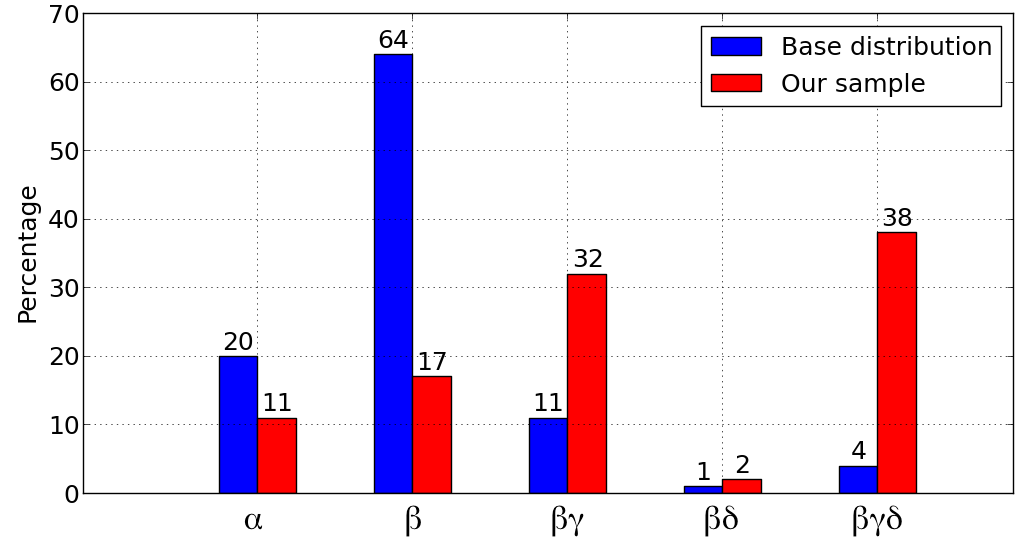}}
   \caption[Distribution of the Mount Wilson classes of the active regions associated with the selected CMEs]{Distribution of the Mount Wilson classes of the active regions associated with the selected CMEs.
   The base distribution is that reported by \citet{jaeggli:16} (calculated over the period 1992-2015).
   }
   \label{fig:4_mountwilson}
   \end{figure}
   
Considering the evolution of sunspot group magnetic classifications, we found that a significant fraction of active regions (16 out of 47) underwent a change in their magnetic topology on the day of the CME eruption, compared to the day before. 
Moreover, 12 of the 16 evolution patterns observed involved changes from or to $\beta \gamma \delta$ configurations. 
   
\medskip
\emph{McIntosh classification.} 
Considering the Modified Z$\ddot{\textup{u}}$rich class of ARs (McIntosh Z code), the most common classes were the \emph{D} and \emph{E} classes (18 and 16 events respectively), followed by \emph{F} topologies (6 events) and \emph{H} topologies (5 events).
The \emph{A} class, corresponding to small unipolar sunspots, is not present in our sample.
Comparing this distribution with the base distribution reported in \citet{mcintosh:90}, 
we find a significant enhancement in the fraction of \emph{F}, \emph{D} and \emph{E} classes in our sample, while \emph{B}, \emph{C} and \emph{H} classes result depressed in their occurrence frequency.

   \begin{figure}
   \centering
   {\includegraphics[width=0.80\columnwidth]{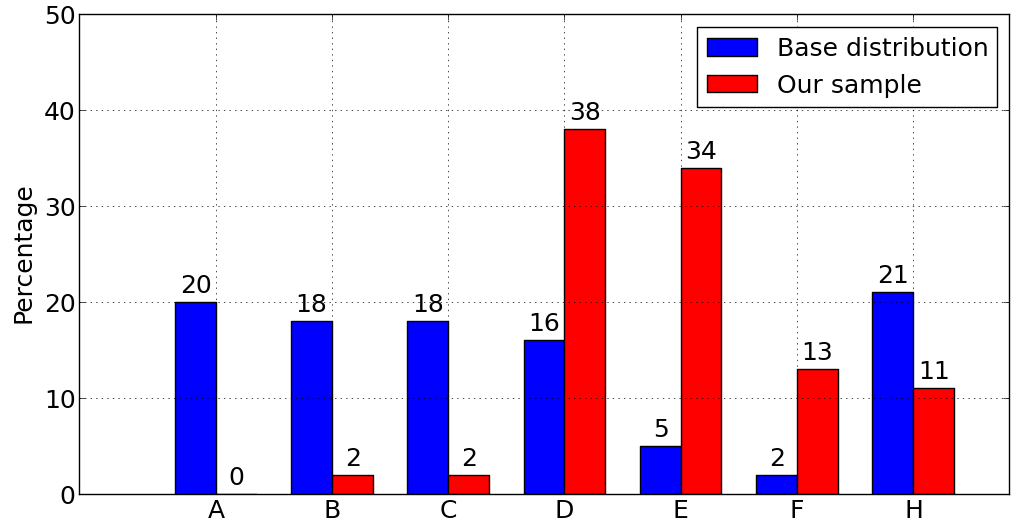}} \\
      {\includegraphics[width=.545\textwidth]{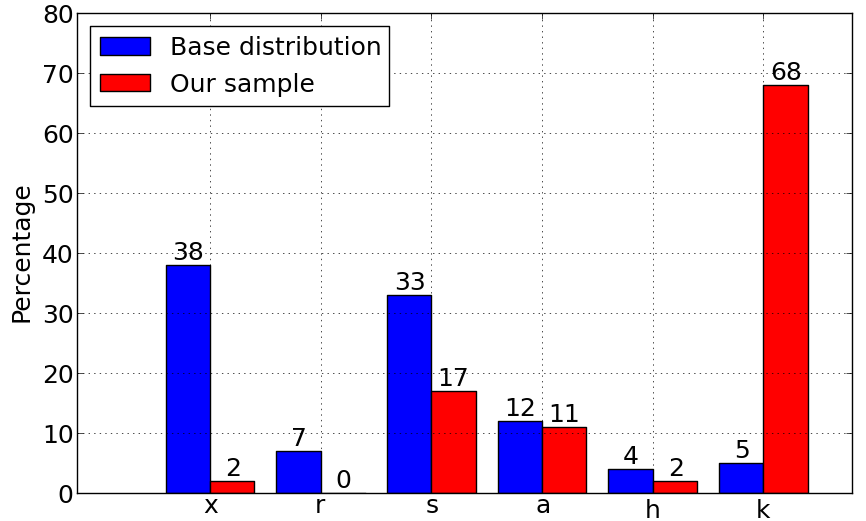}}
   {\includegraphics[width=.445\textwidth]{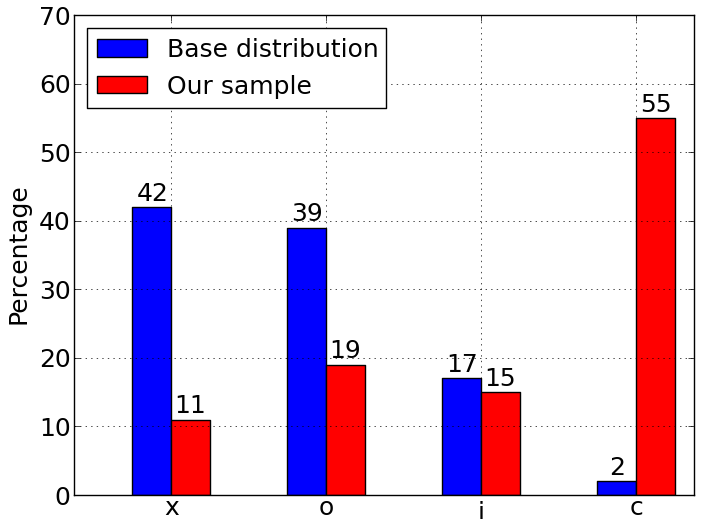}}
\caption[Distribution of the McIntosh classes of the active regions associated with the selected CMEs]
{Distribution of the McIntosh classes of the active regions associated with the selected CMEs. 
\emph{Top}: Z code, \emph{bottom left}: p code, \emph{bottom right}: c code. 
The base distribution is that reported by \citet{mcintosh:90} (calculated over the period 1969-1976).}
   \label{fig:4_mcintosh}
   \end{figure}
   
Turning to penumbral and sunspot distribution classes of ARs (McIntosh p and c codes),
we found that 33 out of 47 active regions ($70$\%) showed a large penumbra (\emph{h} and \emph{k} classes), 
while 32 out of 47 ($68$\%) showed a large penumbra which was asymmetric (classified as \emph{k}).
26 out of 47 regions ($55$\%) showed a compact sunspot distribution, classified as \emph{c}. 
Combining together these two parameters, we obtained the fraction of active regions which are classified as ``complex regions'' by \citet{mcintosh:90}.
We found that 25 out of 47 regions ($53$\%) were classified as \emph{kc} in their combined penumbral/sunspot distribution. 
On average, only $\sim 2$\% of the regions show this degree of complexity.
All these results differ significantly from the base distribution, as reported in Figure~\ref{fig:4_mcintosh}.
In particular, we observe that the presence of complex magnetic topologies in ARs strongly increases the productivity rate of fast CMEs and represents one of the most favourable condition identified in this work's analysis of solar active regions sources of Earth-directed halo CMEs during Solar Cycle 24.
This work's results are in good agreement with those found by \citet{michalek:13} studying a set of 68 ARs associated with halo CMEs observed by LASCO during the years 2001-2004, even though in our sample we observe an even higher tendency for ARs to show \emph{kc} topologies. 
Figure~\ref{fig:4_mcintosh} summarises the above results.
  
\medskip
\emph{CME-productive ARs.}
The overall number of ARs involved is 31, 8 of which gave origin to more than one CME belonging to the sample:
NOAA AR 11261 (Aug. 2011), 11283 (Sep. 2011), 11302 (Sep. 2011), 11402 (Jan. 2012), \ (Mar. 2012), 11944 (Jan. 2014), 12158 (Sep. 2014), 12371 (Jun. 2015). 
These 8 ARs alone produced 23 CMEs out of the 53 events considered.
The most productive one were AR 11429, which gave origin to 6 CMEs (\#18-23) and AR 12371, which gave origin to 4 CMEs (\#49, 51-53).

On average, CME-productive ARs were observed to contain more sunspots than other ARs, and to be larger in terms of their sunspot area.
CME-productive ARs were also extremely flare-productive: all the CMEs originated from this kind of ARs were in fact associated with energetic flares of class $\ge$ M2.0. 
Moreover, 7 out of the 8 X-class flares associated with the selected 52 CMEs for which AR data were available, 
were actually associated to CMEs originated from CME-productive ARs.
On the days associated with the eruptions of the selected CMEs, all these CME-productive regions were classified as \emph{D}, \emph{E} or \emph{F} according to McIntosh classification scheme, and they were all characterised by a large asymmetric penumbra (classified as \emph{k}). 
The only exception was AR 11283 on 6 September 2011, which was classified as \emph{a} (small asymmetric penumbra).
Most of them (6 out of 8) also showed a compact sunspot distribution (classified as \emph{c} in the McIntosh c class).
Overall, 16 out of 23 CMEs originated from CME-productive regions (70\%) were associated with ARs classified as complex (\emph{kc}) regions on the days of the eruptions.

\subsection{Association with solar flares}

Over a set of 53 CMEs under study, 48 of them ($91$\%) were associated with C+ solar flares observed in the Soft X-Ray domain within 2 hours before or after the first CME appearance in the LASCO C2 coronagraph. 39 out of 53 CMEs ($74$\%) were associated with either X- or M-class flares.
This result is in good agreement with previous work results; for example, \citet{wang:02} found an association of $70$\% on a set of 132 frontside (full and partial) halo CMEs in the period 1996-2000.

As shown in Table~\ref{tab:4_flare_class}, among the 48 CMEs associated with flares of importance greater than B, $9$ were associated with X-class flares, $30$ with M-class flares and $9$ with C-class flares.
Interestingly, of the 8 X-class flares of which data concerning the source AR were available, all were associated with \emph{D}, \emph{E} or \emph{F} regions.
Moreover, 7 of them showed a \emph{k} (large asymmetric) penumbral configuration and 6 of them a \emph{c} (compact) spot distribution.
Combining together these parameters, we find that 6 out of 8 X-class flares were originated in complex \emph{kc} active regions.

Finally, considering the duration of the associated flares, we find that 31 out of 48 flares ($65$\%) are impulsive events decaying to half of the peak intensity within 60 minutes from their start, while 17 events ($35$\%) are Long Duration Flares (LDFs). The average duration of flares is 67 minutes, in a range spacing from 247 minutes to 9 minutes.
These results agree with those found in a statistical study of 69 flares observed in association with fast CMEs during Solar Cycle 23 \parencite{lakshmi:13}.

\begin{table}
\centering
	\begin{tabular}{lcc}
	\toprule
	\multicolumn{3}{c}{\textbf{Solar flares: CME association}} \\
	\midrule
	{SXR class}					& {No. of CMEs}			& {Percentage} \\
	\midrule
	X							& $9$						& $17.0$\%		\\
	M							& $30$						& $56.6$\% 		\\
	C							& $9$						& $17.0$\% 		\\
	$\le B$						& $5$						& $9.4$\% 		\\
	\midrule
	Total						& $53$						& $100$\% \\					
	\bottomrule
	\end{tabular}

\smallskip
\caption[Association of CMEs with different classes of SXR flares]{Association of CMEs with different classes of flares. 
Column 3 shows the percentage of CMEs associated to each class with respect to the total number of CMEs.}
\label{tab:4_flare_class}
\end{table}

\subsection{Association with other activity features} 

In this work we have focused on filaments/prominences, X-ray sigmoidal structures and global coronal perturbations with the aim of investigating pre- and post-eruptive conditions at the Sun to fully characterise the large-scale eruptions considered.
Table~\ref{tab:4_other_features} summarises the association rate found in the case of all these additional features.

\begin{table}
\centering
	\begin{tabular}{lcc}
	\toprule
	\multicolumn{3}{c}{\textbf{Other activity features: CME association}} \\
	\midrule
									&No. of CMEs	& Percentage\\
	\midrule
	Filaments						& 19			& 35.8\%\\
	X-ray sigmoid configurations 	& 23			& 43.5\%\\
	Coronal waves					& 38			& 71.7\%\\
	\midrule
	Total							& 53			& 100\% \\
	\bottomrule
	\end{tabular}

\smallskip
\caption[Association of CMEs with additional solar activity features]{Association with additional solar activity features. Column 3 shows the percentage of CMEs associated to each activity feature with respect to the total number of CMEs. Note that the association with filaments includes quiet filaments, eruptive filaments, filament activations, surges and sprays.}
\label{tab:4_other_features}
\end{table}

We observe low association rates compared to previous works, in particular in the case of filaments/prominences.
For comparison, \citet{stcyr:91} found an association of $76$\% with filaments/prominences over a set of 73 CMEs observed by Solar Maximum Mission in the period 1984-86.
One explanation to our low association rates can be traced back to the fact that AR filaments are also often not well defined. 
A second factor could be the fact that, since we are dealing mostly with CMEs originated near the solar disk centre, the identification of eruptive filaments might have been difficult and some eruptions might have passed unnoticed \citep{palmerio:17}. 

Considering X-ray sigmoid configurations, we find a $43$\% association with fast Earth-directed halo CMEs, a result that is much lower than that reported in previous studies. For example, \citet{canfield:99} found a $65$\% association rate comparing flare-productive ARs with X-ray sigmoids by considering a sample of 79 eruptive ARs catalogued by NOAA in the years 1993 and 1997. We suspect that an observational bias may have affected our result by lowering the association rate for close-to-limb CMEs, for which recognition of sigmoidal features is expected to be more difficult due to projection effects.

Large-scale coronal waves appear to have the highest correlation ($>70$\%) with fast CMEs.
This result is in agreement with the association rate found by \citet{cliver:05} for fast Earth-directed CMEs; 
however, in a scenario in which CMEs are expected to trigger global coronal disturbances in most cases, it is difficult to envision the use of coronal waves as indicators of a potentially geoeffective CMEs as this kind of signature would be not distinctive of this class of CMEs only.

\section{Signatures at Earth: comparing simulation results to Wind data}
\label{sec:results_ip}

In order to reconstruct the propagation and evolution of the selected CMEs/ICMEs in the heliosphere up to their arrival at Earth, 
we have compared ENLIL simulation outputs to in-situ measurements obtained by the Wind spacecraft, 
focusing on the identification of ICME-driven forward shocks to assess the performance of our prediction scheme and to investigate the properties of CME transit times.

\medskip
\noindent \emph{Shock identification and transit times.}
To assess the arrival rate of the selected CMEs at Earth, we have applied the forward-shock identification criteria used by the Heliospheric Shock Database developed and maintained at the University of Helsinki (\url{http://ipshocks.fi/}, \cite{kilpua:15}) to 1-minute resolution data relative to magnetic field strength, plasma bulk speed and proton number density from the Wind/MFI and Wind/SWE instruments. In particular, the following criteria have been applied:
\begin{equation}
\label{eqn:shock_b}
\frac{B_ {down}}{B_ {up}} \ge 1.2,
\end{equation}
\begin{equation}
\label{eqn:shock_n}
\frac{N_p^ {down}}{N_p^ {up}} \ge 1.2,
\end{equation}
\begin{equation}
\label{eqn:shock_v}
V_ {down} - V_{up} \ge 20 \,\, \si{\kms},
\end{equation}
where upstream and downstream values were calculated over a fixed time interval $\Delta t _{up}= \Delta t_{down} = 10$ min before and after the shock.
For all the events, we have checked the data to make sure that an identified shock was driven by an ICME and not by other kind of transient events such as Corotating Interaction Regions (CIRs) \citep{jian:06}.
Note that no Wind data were available for the days following the LASCO observation of CMEs \#34, \#46 and \#47,
so we have dropped out these event from the following analysis.

In this case we have uniquely identified a total of 36 forward shocks; 
among those, 5 could be linked to two or more CMEs at the Sun, leading to a total of 41 out of 50 ($82$\%) CMEs driving a forward shock that arrived at Earth.
This result is in agreement with other recent studies on Earth-directed halo CMEs \parencite{shen:14}. 

\begin{figure}[t]
\centering
   {  \includegraphics[width=.95\textwidth]{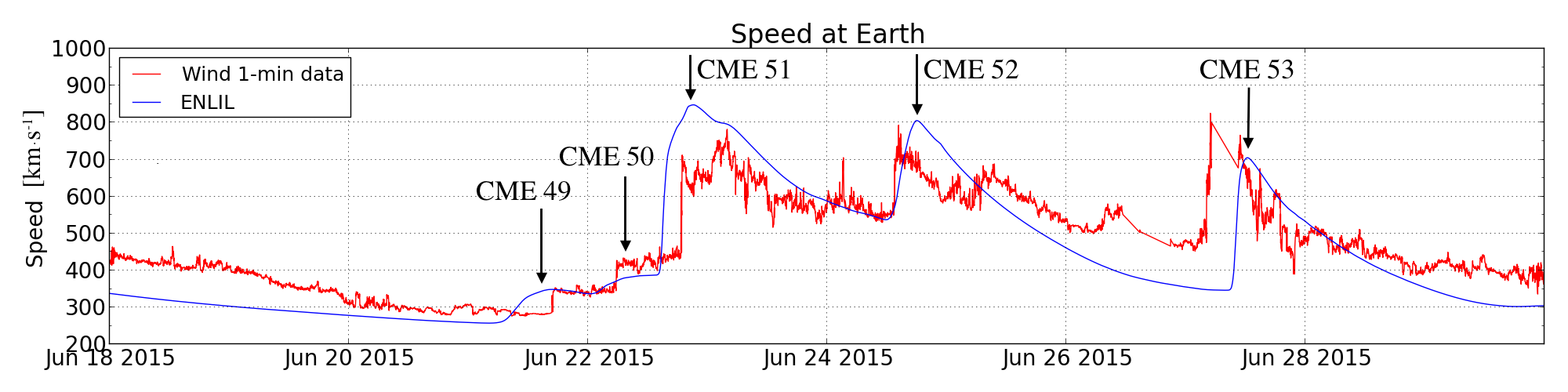} \\
	\includegraphics[width=.95\textwidth]{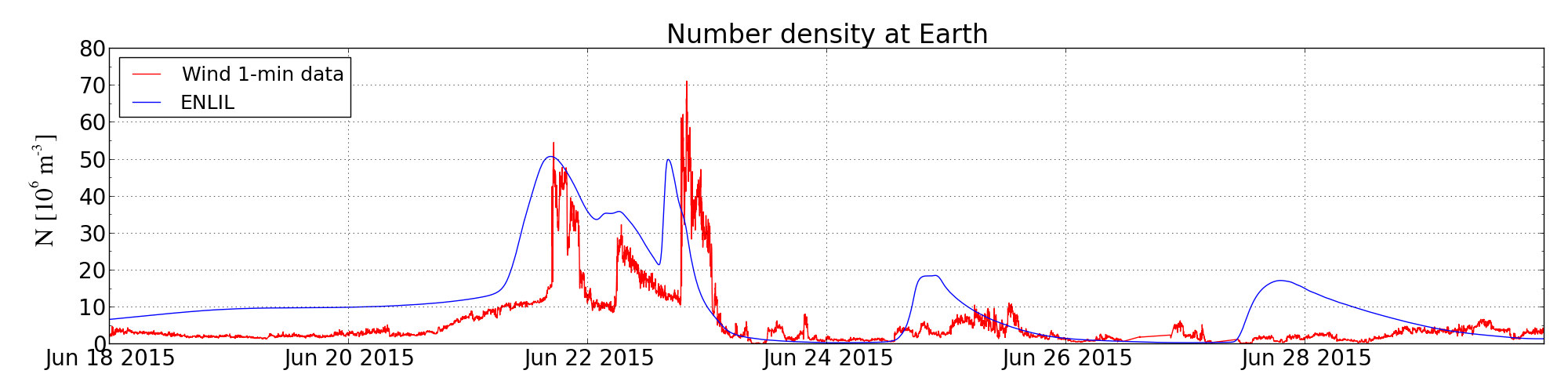} \\
	\includegraphics[width=.95\textwidth]{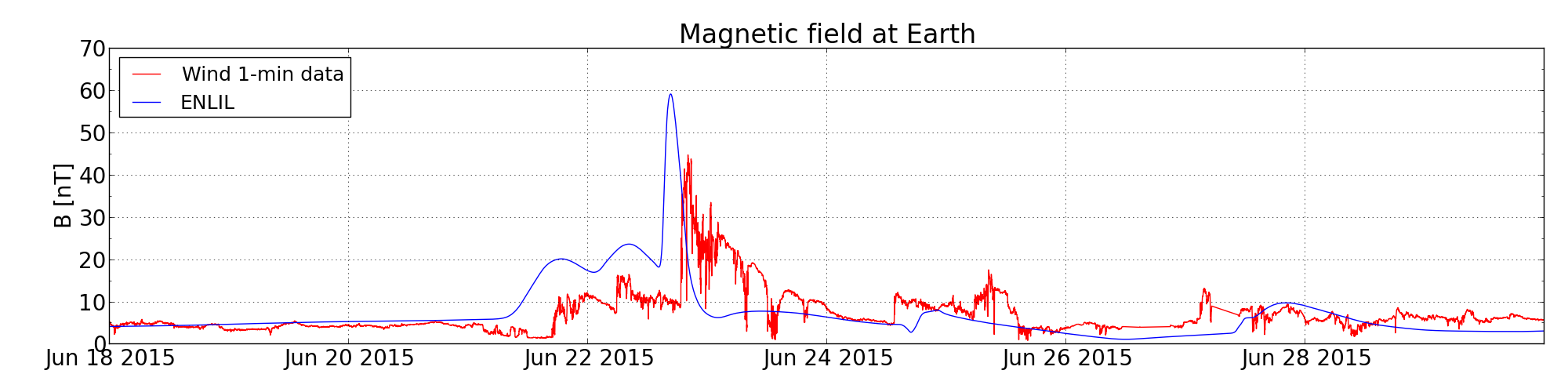} \\
   } 
\caption[Comparison between Wind data and ENLIL simulation results at Earth]{Comparison between Wind data and ENLIL simulation results at Earth for a set of 5 CMEs (\#49-53) observed by LASCO in the days 18-25 June 2015.} 
\label{fig:4_wind_enlil}
\end{figure}

ENLIL time series at a certain position in space are characterised by a variable time step, providing higher time resolution in conjunction with highly variable solar wind conditions. The typical time resolution at Earth for a $256\times30\times90$ grid was between $3$ and $6$ minutes.
To identify forward shocks from ENLIL simulation outputs we have adapted the above shock identification conditions (Equations \ref{eqn:shock_b}, \ref{eqn:shock_n} and \ref{eqn:shock_v}) by considering $\Delta t_{up}$ and $\Delta t_{down}$ as composed of 60 data points each, for a typical $\Delta t \sim 3-6$ hours. 
By applying these conditions to all the ENLIL simulation outputs at Earth, we have been able to uniquely identify a total of 34 forward shocks.
Among the identified shocks, 4 could be linked to multiple CMEs at the Sun that merged within 1 AU, leading to a total of 38 out of 50 ($76$\%) CMEs that were predicted to arrive at Earth.

For all the events that arrived at Earth we computed the \emph{measured transit time},
defined as the time interval between the passage of the ICME forward shock observed by Wind and the estimated CME onset time, 
and the \emph{simulated transit time}, 
defined as the difference between the ENLIL-estimated arrival time of the ICME at Earth and the estimated CME onset time.
The measured transit times range from 34 hours to 76 hours, with an average value of 55 hours and a median of 51 hours.
From simulation outputs we find simulated transit times spacing from 28 hours to 79 hours, with average and median values equal to 51 hours.
Assuming a typical 12-hour interval between the forward shock arrival and the ICME passage, the measured transit times presented here are consistent with those reported in \citet{gopal:01} for a set of 47 CMEs/ICMEs occurred during Solar Cycle 23 in the years 1996-2000.
The mean error, corresponding to the average difference between the measured and forecasted arrival times, calculated over the whole set of events, was of 0.2 hours; 
the median of the error distribution was -0.4 hours. 
The mean absolute error (average the absolute values of the difference between the measured and forecasted arrival times) was 10.8h.
Considering single events, only 52\% of them were predicted to arrive within $\pm 10$ hours from the actual arrival time. 
However, as in this work we are mainly concerned with categorical forecasts, we leave a detailed investigation on the arrival time performance for further studies. 

\medskip
\noindent \emph{CME arrival forecast performance.}
The CME/shock association rate found by analysing ENLIL results is consistent with that observed in Wind data, despite the presence of some misses and false alarms implying that some CMEs that were expected to impact Earth according to simulations, did not actually arrive at Earth location, and vice versa. 
To assess the forecast performance for CME arrivals, from the results of the ICME-driven forward shock analysis we considered the event “CME/shock arrival” condition at Earth position. A ``yes'' forecast was hence marked in-situ by the identification of a forward shock; a ``no'' forecast was associated to a case when the CME did not arrive at Earth or a forward shock was not detected in Wind data.
Considering the contingency tables for the CME arrival prediction, we have
34 correct arrival predictions, 6 correct negatives, 4 false alarms and 6 misses, resulting in a CR rate of 60\% an FA rate of 40\%, a CA ratio of 89\% and an FA ratio of 11\%.
In this case the forecast performed poorly in the case of the FA rate (40\%, with respect to a perfect score =0\%).
On the other hand, the FA ratio score (11\%) was the best among the different forecast scores considered in this work.
Such results indicate that the CME arrival forecast performed particularly well in terms of the number of predicted hits (which dominated the forecast sample), while it gave no useful information in the case of predicted ``no'' events.

\begin{table}
\centering
	\begin{tabular}{lcccc}
	\toprule
	\multicolumn{5}{c}{\textbf{Forecast performances}} \\
	\midrule
	{Forecast type}				& CR rate	& FA rate  	& CA ratio 	& FA ratio \\
	\midrule
	CME arrival					& 60\%		& 40\% 		&89\%		& 11\% \\		
	$K_p$ index (random)		& 75\%		& 25\% 		&80\%		& 20\% \\							
	$K_p$ index (southward)	& 29\%		& 71\% 		&68\%		& 32\% \\
	\bottomrule
	\end{tabular}

\smallskip
\caption[]{Summary of the performances for the CME arrival and $K_p$ index forecasts, including the correct rejection (CR) rate, false alarm (FA) rate, correct alarm (CA) ratio and false alarm (FA) ratio scores defined in Section \ref{sec:prediction_geoeffectiveness}. }
\label{tab:performances}
\end{table}

To identify the cause of the missing detections, we checked the ENLIL movies finding that 3 events (\#11, \#14 and \#15) were predicted as impacting Earth from the flank, while events \#49 and \#50 were predicted to partially merge with event \#51 within 1~AU. This suggests that the reason of the missing detection in ENLIL data may be due to a weak shock signal in the timeseries that led to a failure in the shock detection algorithm applied.
On the other hand, in the case of false alarms we consider the discrepancy as primarily due to the CME width values used as input conditions in our simulations, as not all halo CMEs are characterised by average widths such as those described in Section~\ref{sec:event_selection}, as well as by the errors related to the propagation direction in case of deflected events.
In this sense, a major improvement would come from the analysis of CME stereoscopic images taken by the SECCHI coronagraphs on-board the STEREO spacecraft, as in this case Earth-viewed halo CMEs would appear as quasi-limb events from the STEREO spacecraft and a more precise evaluation of the actual width of single events could be performed (\cite{jang:16} and references therein).
We plan to do a detailed study of the effects that different CME input parameters (width, passage time at 0.1 AU, direction of propagation, etc.) have on the predicted arrival time at Earth and other spacecraft locations in future works. 
Figure~\ref{fig:4_wind_enlil} shows the comparison between Wind data and ENLIL with cone simulation results at Earth for a set of 5 CMEs (\#49-53) observed by LASCO in the days 18-25 June 2015.
The performances of the CME arrival and $K_p$ forecasts are reported in Table \ref{tab:performances}.

\section{Prediction of geoeffectiveness}
\label{sec:results_geo}

\subsection{\texorpdfstring{$K_p$}{K\_p} index}

As shown in Figure~\ref{fig:4_geo}, 34 out of 50 CMEs ($68$\%) for which Wind data were available, 
resulted in a 3-hour $K_p \ge 5$ over the 96 hours after their first appearance in LASCO C2 field-of-view.
30 out of 50 events ($60$\%) triggered a $K_p \ge 6$ storm while 15 out of 50 ($30$\%) triggered a $K_p \ge 7$ storm.
Among the non-geoeffective CME, we found 9 events that were marked in-situ by a shock, but did not trigger any geomagnetic storm 
(CMEs \#11, 17, 21, 32, 36, 37, 44, 49, 53). 
On the other hand, we found 2 cases in which a geomagnetic storm was triggered within 3-4 days from the CME observation 
even though no shock was observed at L1 (CMEs \#24 and 35).
Checking the in-situ data, we found CIR signatures before the storm onsets, meaning that in such cases the storms were not triggered by CMEs.
We therefore classified CMEs \#24 and 35 as non-geoeffective. 

By considering $K_p =5$ as minimum geoeffectiveness threshold, about one-fourth of the CMEs has not resulted in a significant $K_p$ impact.
A fraction of $68$\% geoeffective CMEs among all the selected Earth-directed halo CMEs is significantly higher than that found, for example, by \citet{wang:02} using the same geoeffectiveness criteria. 
They reported that only $45$\% of Earth-directed halo CMEs are geoeffective, by considering a set of 132 frontside (full and partial) halo CMEs in the period 1996-2000.
This discrepancy is reasonably due to the fact that in this work we have considered only fast CMEs and it confirms, \emph{a posteriori}, our choice of fast CMEs as the most geoeffective type.
%
As reported in Figure~\ref{fig:4_geo}, the impact on Earth in terms of the $K_p$ varies with the set of events and the $K_p$ threshold considered.
The association with K$_p \ge 5$ was $63-78$\% depending on the set considered, with the highest association rate ($78$\%) for events belonging to S2. 
This result confirms the existence of an ``east-west asymmetry'' in the source locations of geoeffective CMEs, with regions close to the disk centre and in the western hemisphere being the most favourable sources of potentially geoeffective events (\cite{michalek:06}).
The association with K$_p \ge 6$ was $44-68$\% depending on the set considered, with the highest association rate ($68$\%) for events belonging to S3. 
The association with $K_p \ge 7$ storms was lower than $35$\% for all subsets.
%
Finally, the inverse-checking procedure allowed to associate the majority of geomagnetic storms (K$_p \ge 5$) of Cycle 24 with Earth-directed halo CMEs observed at the Sun. 
However, a significant fraction of geomagnetic storms showed no temporal association with any halo CME observed by LASCO coronagraphs. 
In such cases, the source of the observed geomagnetic activity is expected to be a partial-halo CME or a CIR.

\begin{figure}
\centering
   {\includegraphics[width=.80\textwidth]{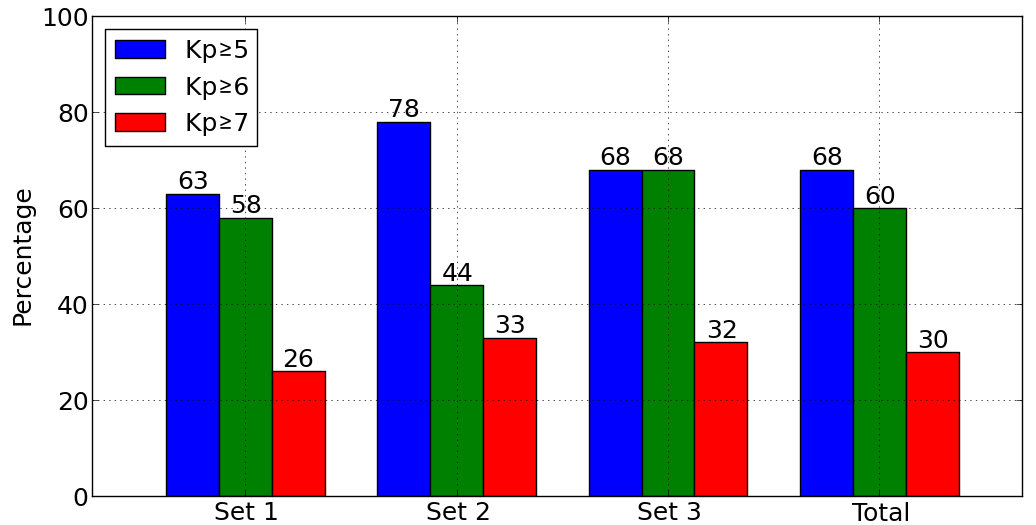}}
\caption[Fraction of CMEs associated with $K_p \ge 5$, $\ge 6$ and $\ge 7$ over the next 96 hours]{Fraction of CMEs that triggered a $K_p$ geomagnetic storm over the 96 hours after their first appearance in LASCO C2 coronagraph, by set.}
\label{fig:4_geo}
\end{figure}

\medskip
\noindent \emph{$K_p$ predictions}.
Figure~\ref{fig:4_geo_kp} reports the measured and predicted number of storms associated with CME/ICME pairs, by $K_p$ intensity.
The predicted $K_p$ indices are reported in the case of both southward and random orientations.
As expected, completely southward $\bm{B}$ fields would lead to much stronger effects on Earth, due to the enhanced magnetic reconnection rate triggered at dayside magnetopause. 
Overall, this worst-case scenario tends to overestimate the maximum $K_p$ value in most of the events; 
the number of events associated with a $K_p \le 7$ is significantly depressed, 
while that of the events associated with a $K_p \ge 8 $ is greatly enhanced (23 $K_p \ge 8$ events compared to the 9 observed events).
However, the estimated number of 39 events resulting in a $K_p \ge 5$ is very close to the actually observed number of 36 events.
Using the expectation value for the clock-angle term (random case), the predicted maximum $K_p$ distribution closely resembles actual measurements in the number of highly geoeffective ($K_p \ge 8$) events. 
On the other hand, it highly overestimates the number of non-geoeffective ($K_p \le 4$) events.
Such results emphasise the crucial importance of developing models capable to reliably predict the $\bm{B}$ orientation of Earth-impacting ICME structures, as cone CME models and artificial $\bm{B}$ orientations such as those considered in this work cannot fully capture the intensity distribution of geomagnetic storms.

\begin{figure}
\centering
   {\includegraphics[width=.82\textwidth]{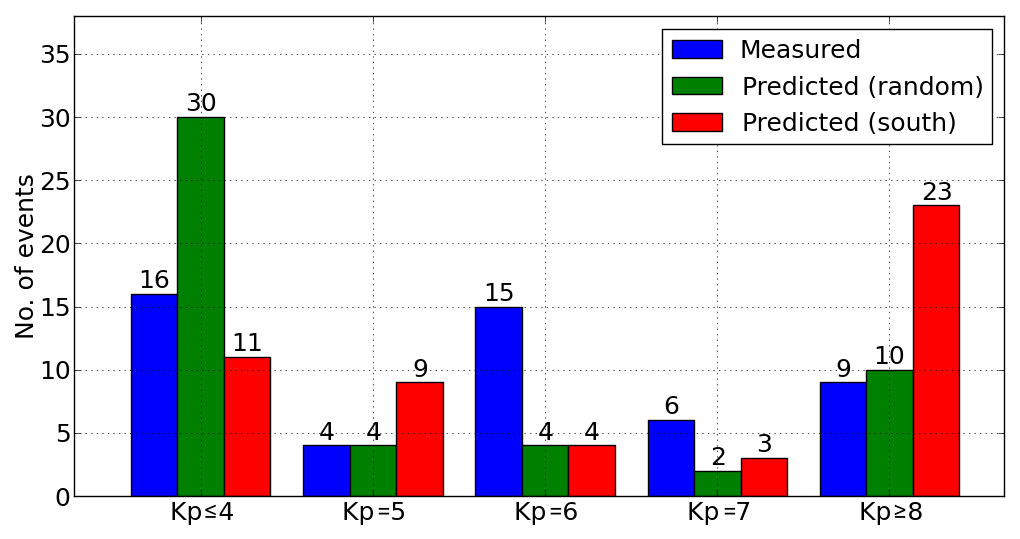}}
\caption[]{Comparison between the measured and predicted geomagnetic storm intensities, based on maximum $K_p$ index. 
}
\label{fig:4_geo_kp}
\end{figure}


\medskip
\noindent \emph{$K_p$ forecast performance.} %
To test the $K_p$ index prediction performance, we have considered the ``event'' to be the ``geomagnetic storm'' defined by a $K_p \ge 5$ condition. 
A ``yes'' forecast was then labelled by a $K_p \ge 5$ prediction, while a ``no'' condition was associated to a $K_p \le 4$.
The contingency table relative to the average-case scenario of a randomly orientated IMF was characterised by a total of
16 hits, 18 misses, 4 false alarms and 12 correct negatives, leading to a CR rate of 75\%, an FA rate of 25\%, 
a CA ratio of 80\% and an FA ratio of 20\% (see Table~\ref{tab:performances}).
In this case the forecast performed generally well, with good scores for both the FA ratio and the FA rate.
As observable in Table~\ref{tab:performances}, the FA ratio for this particular forecast was the second best predicted one among all the different forecasts considered in this work. However, in this case there was a high number of misses, \emph{i.e.} many events that arrived at Earth were not predicted to impact.
For the worst-case scenario of a totally southward oriented IMF the contingency table gave 
26 hits, 7 misses, 12 false alarms and 5 correct negatives, giving
a CR rate of 29\%, an FA rate of 71\%, 
a CA ratio of 68\% and an FA ratio of 32\%.
In this case the forecast performed poorly especially in terms of the number of false alarms (FA rate = 71\%, with respect to a perfect score = 0\%), meaning that in this case the forecast significantly overestimated the number of events that arrived at Earth.
On the other hand, the FA ratio score was still good (32\%) although not as good as in the randomly-oriented case.
The number of misses in this case was reduced by a factor of 2 with respect to the previous case.
In conclusion, the forecast performance analysis for the $K_p$ index indicates that the randomly-oriented $K_p$ forecast performed better than the completely southwardly-oriented one in the case of false alarms, but it performed worse in the case of misses. 

As a final note, we point out the case of CME \#37, which was a false alarm in our prediction scheme for both the clock angle orientations considered.
ENLIL in this case did not forecast correctly the event, predicting a face-on impact on Earth and a strong geomagnetic storm ($K_p=9$) for both the orientations considered.
A complete event analysis was presented by \citet{moestl:15}, who attributed the unexpected low geomagnetic impact of the event as due to a non-radial deflection away from the Sun-Earth line to the west due to a nearby coronal hole on the east side of the CME source region. 
Such ``channelling'' effect has been observed to act $<21.5 \,\, R_\odot$ from the Sun, hence within the ENLIL inner heliospheric boundary.
In this sense, we believe that ENLIL missed this channelling effect due to the fact that CMEs are assumed to have a totally-radial motion during insertion in the heliosphere at $<21.5 \,\, R_\odot$. 
Therefore, by taking as input condition the position of the associated AR for the CME direction of propagation, we ignored all the deflection effects that may have acted on the CME between the photosphere and $21.5 \,\, R_\odot$. To try to balance out these effects one could re-calculate the CME propagation direction at $21.5 \,\, R_\odot$ by using more sophisticated reconstruction methods such as the forward-modelling technique \citep{thernisien:09}.

\subsection{Magnetopause stand-off distance}

When considering the geoeffectiveness of CME/ICME pairs in terms of the induced magnetospheric compression, no continuous monitor is available for comparison.
On average, the magnetopause stand-off distance along the Sun-Earth direction is $\sim 10 \, R_\textup{E}$ for unperturbed solar wind conditions.
By applying Equation~\ref{eqn:magnetopause_standoff} to ENLIL-at-Earth data series, we have found an average and median values of the distribution of the minimum magnetopause stand-off distance of $5.3 \, R_\textup{E}$, with a maximum compression of $4.4 \, R_\textup{E}$.

According to our estimates, the compression driven by the identified interplanetary forward shocks is sufficient to push the magnetopause within the region of geosynchronous orbits, located at about $6.6 \, R_\textup{E}$ from the Earth centre, in the case of 41 out of 50 CMEs (82\%).
The minimum magnetopause stand-off distance gives an idea of the geoeffectiveness of CME/ICME events in terms of their hazard to satellite operations,
as spacecraft designed to survive in a region of space normally shielded by the Earth's magnetic field become subjected to a much harsher environment once they have entered the magnetosheath.
Moreover, 17 events (43\% of the ENLIL-identified forward shocks) are expect to compress the magnetopause at altitudes $\le 4.9 \, R_\textup{E}$, causing disturbances to navigation systems such as GPS satellites orbiting in MEO at $\sim 26,000$ \si{\km} above the Earth surface.

Figure~\ref{fig:4_11429_earth} shows a comparison of the predicted and measured geoeffectiveness caused by the impact of CMEs \#18-23 in March 2012.

\begin{figure}
\centering
   {\includegraphics[width=.95\textwidth]{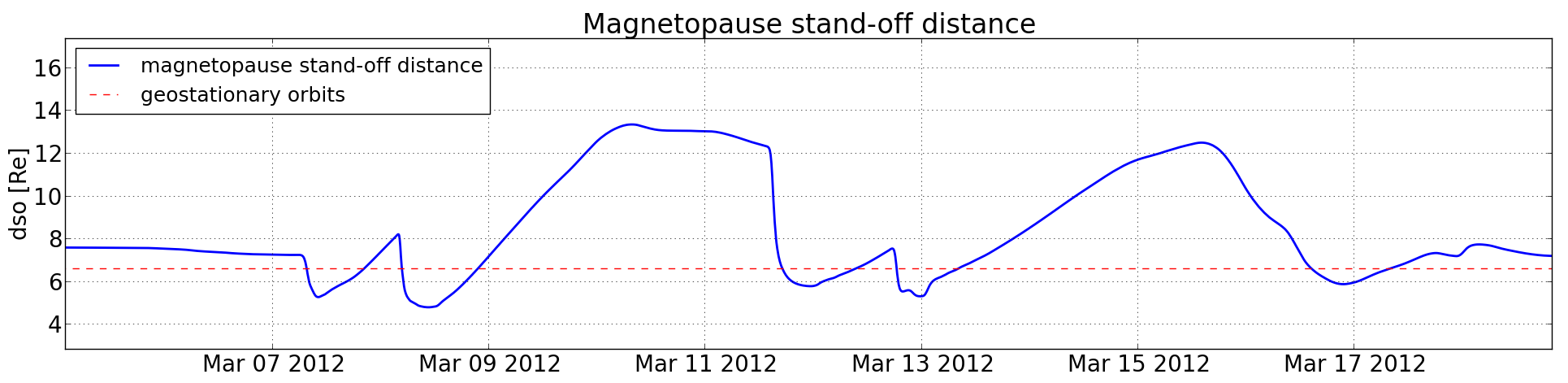} \\
	\includegraphics[width=.95\textwidth]{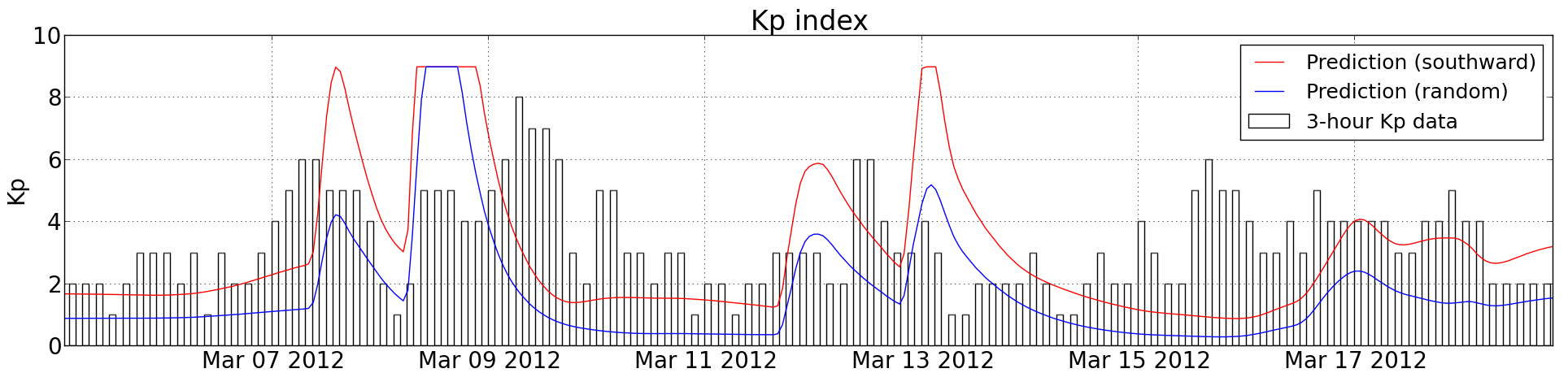}
}
\caption[]{Predicted and observed effects at Earth for geomagnetic storm of 5-13 March 2012, caused by CMEs \#18-23. \emph{Top}: magnetopause stand-off distance. \emph{Bottom}: 3-hour $K_p$ index measured on ground (bars) and predicted from ENLIL outputs at Earth (red and blue). 
}
\label{fig:4_11429_earth}
\end{figure}

\section{Association with major SEP events}
\label{sec:sep}

In order to fully characterise the impact of the selected CME events on geospace, 
we have analysed their association with major SEP events (peak proton particle flux above 10 \si{\mega\eV} $>10$ pfu),
starting from the SEP event lists maintained 
by the NASA CDAW Data Center (\url{http://cdaw.gsfc.nasa.gov/CME_list/sepe/}) 
and by NOAA (\url{http://umbra.nascom.nasa.gov/SEP/}).
In the years 2009-2015, a total of 41 major SEP events have been observed 
and the majority of them (92\%) was caused by a halo CME.
Among those 41 major SEP events, 19 (44\%) were associated to the CMEs included in our analysis.
Addressing the problem the other way around, we found that 19 of the 53 CMEs considered in the analysis (36\%) triggered a major SEP within few hours from their onset, 
and 4 of them were extremely intense events associated with a peak proton particle flux above 10 \si{\mega\eV} $>1,000$ pfu, three orders of magnitude above the minimum threshold for major SEPs.

Considering the source locations of SEP-associated CMEs, we observed a strong ``east-west asymmetry'' in their heliographic distribution, 
with the majority of them originated in the west part of the visible solar disk.
In fact, the 13 out of 19 SEP-associated CMEs (68\%) originated from the western hemisphere of the Sun, with an average source solar longitude equal to 20$^\circ$ W.
For comparison, the average solar longitude for SEP-less CMEs was equal to 3$^\circ$ E, while considering all the CMEs regardless of their association with SEP events, the average source longitude was 5$^\circ$ W.

Investigating the association of SEP-CME events with solar flares, we found that all the 19 SEP-associated CMEs were also associated with a C+ flare (C:16\%-M:52\%-X:32\%).
For comparison, in the case of the 34 SEP-less CMEs, the flare class distribution was no/B-class flare:15\%-C:18\%-M:59\%-X:9\%.
The flare class distribution for the whole CME set, regardless of their association with major SEPs, was no/B-class flare:10\%-C:17\%-M:57\%-X:17\%.
This result suggests that strong (X-class) flares are observed  more often than usual in the case of CMEs triggering major SEPs 
(32\% compared to 9\% for SEP-less CMEs), but C-class flares account for a significant fraction of the total distribution in both cases (16\% compared to 30\% for SEP-less CMEs).
In our set, association of SEP-CME events with large (X or M) flares is 84\%, slightly higher than that reported by \citet{gopal:03} for a set of 48 major SEP events associated to CMEs occurred during Solar Cycle 23 (1997-2001).
The correlation between the flare class and the intensity of the associated SEP event seems to be very weak (0.36), comparable to that reported by \citet{gopal:03}.

Considering the speed properties of SEP-associated CMEs, they were found to be significantly faster than SEP-less ones.
They were characterised by an average speed in space equal to $1692$ \si{\kms}, while SEP-less CMEs had an average speed of $1162$ \si{\kms}.
Correlating the CME speed to the intensity of the associated SEP-CME events, we find a correlation coefficient of 0.69, slightly higher than that reported by previous studies
(\citealt{kahler:01}; \citealt{gopal:03}).

Finally, $74\%$ of the CMEs that triggered a major SEP event were found to be geoeffective in terms of magnetopause compression and/or $K_p$ index.
For comparison, $71\%$ of the SEP-less CMEs and $72\%$ of all the CMEs considered (regardless of their association with major SEPs) 
were geoeffective according to at least one of the geoeffectiveness conditions considered. 
Figure~\ref{fig:4_kp_sep} shows a comparative plot of the CME events, reporting their association with major SEP events and their maximum $K_p$ index, with a color code for the $K_p$ bar based on the CME speed.
Such results show that while the CME SEP-production efficiency correlates very well with the CME speed, 
the CME geoeffectiveness is the result of the interplay of multiple factors. 

\begin{figure}
\centering
   {\includegraphics[width=.98\textwidth]{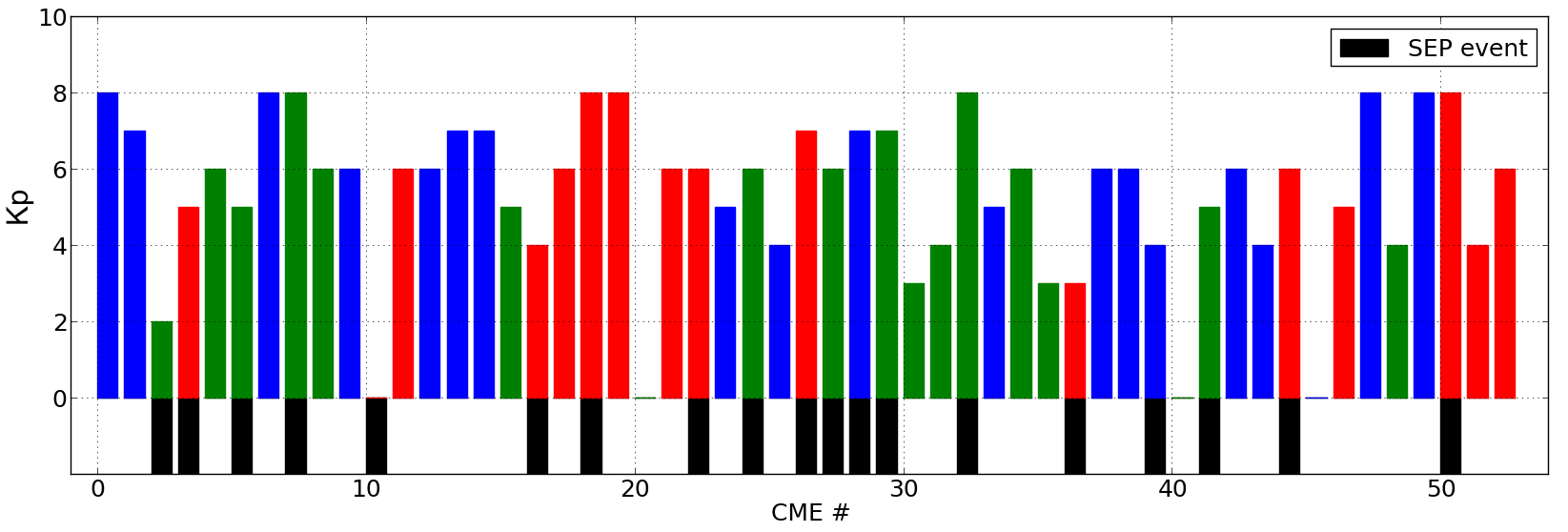}}
\caption[]{Comparative plot of the CME events, showing their association with major SEP events (black bars) compared to their maximum $K_p$ index. The color code for the $K_p$ bar is based on the CME speed: blue for $V_{space} < 1100$  \si{\kms},  green for 1100 \si{\kms} $\le V_{space} < 1500$  \si{\kms}, and red for $V_{space} \ge 1500$  \si{\kms}.}
\label{fig:4_kp_sep}
\end{figure}

Being aware of the importance of CME-CME interactions for particle acceleration \parencite{gopal:02}, 
we have paid particular attention to CMEs that originated from the same AR within a period of some days. 
As seen in Table~\ref{tab:events}, among the 53 CMEs under study, 23 were originated from CME-productive regions, for a total of 8 ARs involved.
Among these 8 sequences of CMEs, 7 (88\%) triggered a major SEP event in coincidence of one of the CMEs involved.
The only exception was represented by CMEs \#9 and \#10, which were launched within a 20-hour interval but in which case the first CME was faster than the second one.
Considering the remaining 4 double-CME events, the SEP peak flux was observed to be coincident with the launch of the second CME in all cases:
this result supports the idea that the particle acceleration efficiency increases as consequence of CME-CME interactions and that the most probable configuration is the case of a fast CME that reaches up a slower CME launched a few hours before from the same solar region. 


\section{Conclusions}
\label{sec:conclusions}

In this work we have presented a statistical analysis of a set of 53 fast ($V \ge 1,000$ \si{\kms}) Earth-directed full halo CMEs 
observed by the SOHO/LASCO instrument during the period Jan. 2009-Sep. 2015, 
and we have then used this CME sample to test the forecasting capabilities of a Sun-to-Earth prediction scheme - based on 3D simulations and solar wind-geomagnetic activity coupling functions - for the geoeffectiveness of Earth-directed halo CMEs.

We have first analysed the solar conditions associated with each CME by considering their association with other solar activity features such as active regions, solar flares, filaments/prominences, X-ray sigmoids and global coronal disturbances, with the final aim of identifying recurrent peculiar features that can be used as precursors of CME-driven geomagnetic storms.
We have found that the solar regions that most likely generate geoeffective CMEs are active regions showing bipolar magnetic topologies with large asymmetric penumbrae and a compact sunspot distribution - usually referred to as ``complex'' topologies (Figures~\ref{fig:4_mountwilson} and \ref{fig:4_mcintosh}).
The presence of energetic flares, in particular  X- and M-class ones, is also a highly favouring factor (Table~\ref{tab:4_flare_class}).
Among the other activity features considered, coronal waves have been the only ones observed in more than one half of the events (Table~\ref{tab:4_other_features}).
Moreover, this study reports a high level of correlation between CME speed and source location properties and the resulting geoeffectiveness, with fast events originating from the central and western regions of the solar disk having the highest probability of resulting in a geomagnetic storm (Figure~\ref{fig:4_geo}).

To reconstruct the global evolution into interplanetary space up to Earth location, 
we have propagated all the CME events into the heliosphere up to 2 AU by means of the WSA-ENLIL+Cone model running at NASA/CCMC.
From the comparison of ENLIL-at-Earth results with in-situ solar wind measurements at L1 obtained by the Wind spacecraft, we have been able to link ICME-driven interplanetary forward shocks to the majority of CMEs, confirming the fact that Earth-directed halo CMEs or their associated shocks do arrive at Earth in most cases. 
From Wind data series we were able to uniquely identify IP forward shocks for $82\%$ of the selected CMEs, and a similar association rate was found in the case of ENLIL-at-Earth results (76\%). Over the totality of the simulated events, the majority (78\%) were correctly forecasted. Among the 11 incorrectly forecasted events, a further analysis showed that most of these events in simulations were predicted to impact Earth from the flank, leading to a weak shock signature in ENLIL timeseries that passed unnoticed by the detection algorithm. The CME input conditions used may have also played a role in limiting the forecast accuracy.

\medskip
In the last part of our analysis we have used simulation outputs upstream of Earth to predict the geoeffectiveness of each CME event 
in terms of the expected geomagnetic activity level and magnetospheric compression.
Our prediction of the induced geomagnetic activity in terms of the 3-hour planetary $K_p$ index was primarily affected by the unreliable ENLIL prediction of the ICME magnetic field orientation at L1. For this reason, we envisioned a worst-case scenario with a completely southward IMF and an average-case scenario of a randomly-oriented IMF.
We found that the worst-case scenario tends to overestimate the single-event $K_p$ value, 
but it generally well represents the fraction of $K_p \ge 5$ events over the total (78\%, compared to 68\% for actual measurements).
On the other hand, a random orientation of the $\bm{B}$ field heavily underestimates the fraction of CMEs causing a mild geomagnetic activity, while it well reproduces the number of strong storms (20\%, compared to 18\% for actual measurements) (Figure~\ref{fig:4_geo_kp}).
Improvements on this point would come from the use of an MHD model that includes the CME internal magnetic structure \emph{e.g.} modelled as a flux-rope CME.
Such kind of models are under development \citep{shiota:16} but none is currently available for extensive modelling such as that needed in the case of the statistical analysis presented in this work.

Our predictions of the minimum magnetopause stand-off distance suggest that the dayside magnetopause is compressed below the altitude of geosynchronous (GEO) satellites whenever one of the identified ICMEs impacts on Earth. In fact, in our study, all the CMEs that arrived at Earth compressed the magnetopause below GEO orbits.
In a significant number of cases, it was compressed even below MEO orbits (42\% of the CMEs that arrived at Earth). 
This result confirms the importance of studying fast Earth-directed CMEs as they can severely affect human activities by generating major disruptions to satellites serving as telecommunication and navigation systems. 

The analysis of related SEP events shows that 74\% of the CMEs associated with major SEPs were geoeffective \emph{i.e.} triggered a minor to intense geomagnetic storm ($K_p \ge 5$). Analysing the association of the selected CMEs with major SEP events, we found that CMEs characterised by higher speeds and originating close to the solar disk centre or from the western hemisphere are much more likely to trigger a major SEP event at Earth. Moreover, the SEP production resulted enhanced in the case of fast CMEs, with a correlation coefficient between CME speeds and SEP peak fluxes of 0.69. Under these conditions, we found that the likelihood for a multiple-CME event to originate a strong SEP as consequence of CME-CME interactions was considerably high.

\medskip
The results of our prediction scheme appear promising as the forecast performances for the CME arrival and $K_p$ index in terms of the scores reported in Table~\ref{tab:performances} show good agreement with in-situ observations and actual data records for geomagnetic activity; improvements to this prediction scheme could come from an extension of the CME sample considered. 
Moreover, as already pointed out, the internal magnetic structure of ICMEs represents a critical issue in determining their impact on geospace and future models will have to take this point into account to provide more reliable space weather predictions.
Finally, in the case of halo CME simulations, a big improvement 
would be represented by the use of stereoscopic images taken by SECCHI coronagraphs on-board the STEREO spacecraft when they are in quadrature with the Earth to determine the kinematical CME properties and hence provide better input parameters for heliospheric CME models; 
in this case, Earth-viewed halo CMEs would appear as limb CMEs and a more precise evaluation of the actual width and speed and direction of propagation could be performed.
In this sense, the prediction scheme presented in this work and the open questions arisen from our analysis represent a promising starting point for future studies.

\begin{acknowledgements}
Part of this work was carried out at the University of Trieste, Department of Physics.
C. S. was partially funded by the Research Foundation - Flanders (FWO) SB PhD fellowship number 1S42817N.
The paper uses data from the SOHO/LASCO CME catalogue, generated and maintained at the CDAW Data Center by NASA and the Catholic University of America in cooperation with the Naval Research Laboratory. SOHO is a project of international collaboration between ESA and NASA.
The Wind solar wind plasma and magnetic field data were obtained at NASA's CDAWeb (\url{cdaweb.gsfc.nasa.gov}). 
The Dst geomagnetic index data were obtained from the World Data Center for Geomagnetism in Kyoto, Japan. 
Final definitive $K_p$ indices are from the Helmholtz Center Potsdam GFZ German Research Centre for Geosciences. 
Simulation results have been provided by the Community Coordinated Modeling Center at Goddard Space Flight Center through their public Runs on Request system (\url{http://ccmc.gsfc.nasa.gov}). The CCMC is a multi-agency partnership between NASA, AFMC, AFOSR, AFRL, AFWA, NOAA, NSF and ONR. The WSA-ENLIL+Cone Model was developed by the D. Odstrcil at the Univ. of Colorado at Boulder.
The authors thank two anonymous referees for constructive comments and suggestions. The editor thanks two anonymous referees for their assistance in evaluating this paper.
\end{acknowledgements}


\printbibliography


\end{document}